\documentclass[journal=jacsat,manuscript=article]{achemso}

\usepackage{chemformula} 
\usepackage[T1]{fontenc} 

\usepackage{amsmath,amssymb,amsfonts}%
\usepackage{amsthm}%

\author{Nian Wu}
\email{nian.wu@aalto.fi}
\affiliation{Department of Applied Physics, Aalto University, Helsinki, 02150, Finland.}

\author{Markus Aapro}
\affiliation{Department of Applied Physics, Aalto University, Helsinki, 02150, Finland.}

\author{Joakim S. Jestilä}
\affiliation{Department of Applied Physics, Aalto University, Helsinki, 02150, Finland.}

\author{Robert Drost}
\affiliation{Department of Applied Physics, Aalto University, Helsinki, 02150, Finland.}

\author{Miguel Martínez García}
\affiliation[Fifth University]{Departamento de Química Orgánica, Universidad Autónoma de Madrid, Madrid, 28049, Spain}
\alsoaffiliation[Seventh University]{IMDEA-Nanociencia, Campus de Cantoblanco, Madrid, 28049, Spain}

\author{Tomás Torres}
\affiliation[Fifth University]{Departamento de Química Orgánica, Universidad Autónoma de Madrid, Madrid, 28049, Spain}
\alsoaffiliation[Sixth University]{Institute for Advanced Research in Chemical Sciences, Universidad Autónoma de Madrid, Madrid, 28049, Spain}
\alsoaffiliation[Seventh University]{IMDEA-Nanociencia, Campus de Cantoblanco, Madrid, 28049, Spain}

\author{Feifei Xiang}
\affiliation{nanotech@surfaces Laboratory, Empa-Swiss Federal Laboratories for Materials Science and Technology, Dübendorf, 8600, Switzerland}

\author{Nan Cao}
\affiliation{Department of Applied Physics, Aalto University, Helsinki, 02150, Finland.}

\author{Zhijie He}
\affiliation{Department of Computer Science, Aalto University, Helsinki, 02150, Finland.}

\author{Giovanni Bottari}
\affiliation[Fifth University]{Departamento de Química Orgánica, Universidad Autónoma de Madrid, Madrid, 28049, Spain}
\alsoaffiliation[Sixth University]{Institute for Advanced Research in Chemical Sciences, Universidad Autónoma de Madrid, Madrid, 28049, Spain}
\alsoaffiliation[Seventh University]{IMDEA-Nanociencia, Campus de Cantoblanco, Madrid, 28049, Spain}

\author{Peter Liljeroth}
\email{peter.liljeroth@aalto.fi}
\affiliation{Department of Applied Physics, Aalto University, Helsinki, 02150, Finland.}

\author{Adam S. Foster}
\email{adam.foster@aalto.fi}
\affiliation{Department of Applied Physics, Aalto University, Helsinki, 02150, Finland.}
\alsoaffiliation[Second University]{WPI Nano Life Science Institute, Kanazawa University, Kanazawa, 610101, Japan.}

\title[An \textsf{achemso} demo]
  {Precise large-scale chemical transformations on surfaces: deep learning meets scanning probe microscopy with interpretability}

\keywords{Automation, tip-induced chemical reaction, machine learning, reinforcement learning, Br removal, STM, reaction mechanism, DFT, BOSS}

\begin{document}


\begin{tocentry}

AutoOSS automates tip-induced Br removal from large-scale ZnBr2Me4DPP molecules on Au(111) in scanning tunneling microscopy with interpretability and selectivity by combining theoretical calculations and advanced machine learning techniques.

\end{tocentry}

\begin{abstract}
  Scanning Probe Microscopy (SPM) techniques have shown great potential in fabricating nanoscale structures endowed with exotic quantum properties achieved through various manipulations of atoms and molecules. However, precise control requires extensive domain knowledge, which is not necessarily transferable to new systems and cannot be readily extended to large-scale operations. Therefore, efficient and autonomous SPM techniques are needed to learn optimal strategies for new systems, in particular for the challenge of controlling chemical reactions and hence offering a route to precise atomic and molecular construction. In this paper, we developed a software infrastructure named AutoOSS (\textbf{Auto}nomous \textbf{O}n-\textbf{S}urface \textbf{S}ynthesis) to automate bromine removal from hundreds of Zn(II)-5,15-bis(4-bromo-2,6-dimethylphenyl)porphyrin (\ch{ZnBr2Me4DPP}) on Au(111), using neural network models to interpret STM outputs and deep reinforcement learning models to optimize manipulation parameters. This is further supported by Bayesian Optimization Structure Search (BOSS) and Density Functional Theory (DFT) computations to explore 3D structures and reaction mechanisms based on STM images.
\end{abstract}

\section{Introduction}
Precisely and controllably manipulating atoms or molecules on surfaces offers the potential for assembling nanomaterials with tuneable exotic properties for novel applications in optoelectronics and spintronics\cite{khajetoorians2019creating, pitters2024atomically, gross2018atomic, sanvito2011molecular, li2022manipulating, li2021recent, ren2021plasmonic}. Recently, scanning probing microscopy (SPM), including scanning tunnelling microscopy (STM) and atomic force microscopy (AFM), has shown great potential in nanofabrication through complex manipulations including pulling, pushing, pick–transfer–drop and dissociation\cite{whitman1991manipulation, zhang2023nanoscale, li2023research, simpson2019control,zhong2023surface}. These manipulations are predominantly controlled through the tip position (tip$_x$, tip$_y$, tip$_z$), bias voltage (V) and tunnelling current (I) in STM. However, the selection and optimization of such parameters is a time-consuming and repetitive process and strongly depends on the domain knowledge, which is not necessarily transferable to new systems. Therefore, efficient and autonomous SPM techniques are needed to reduce the reliance on human supervision and efficiently learn optimal strategies for the fabrication of functional nanostructures, particularly to the scale that would have an impact in real technologies.

Advanced machine learning techniques, especially image classification, image segmentation and reinforcement learning (RL), have recently emerged as promising methods to automate various tasks in SPM, including the identification of optimal sample regions, the evaluation of the quality of scanning images, tip conditioning and the selection of manipulation parameters and the detection of reaction sites \cite{kalinin2021automated,rashidi2018autonomous, krull2020artificial, alldritt2022automated, su2024intelligent, zhu2024autonomous, ramsauer2024mam}. For example, RL decision-making agents have been developed using discrete actions to find the proper trajectories to lift a large molecule \cite{leinen2020autonomous}, and also to laterally manipulate a polar molecule \cite{ramsauer2023autonomous}. In contrast to making decisions within a set of discrete actions, Chen and coworkers developed a Deep Reinforcement Learning (DRL) approach capable of selecting parameters from continuous action space including tip-start and -end positions, bias voltage and tunnelling conductance to steer the motion of atoms\cite{chen2022precise}. The advancements in SPM automation mentioned above pave the way for the next step in nanostructure assembly -- the automation of chemical reactions. 

For the engineering of new organic materials, on-surface synthesis (OSS), which is based on chemical reactions, has developed into a powerful tool for the controllable formation of molecular structures on solid surfaces \cite{clair2019controlling}. In particular, the ability to control chemical reactions using temperature \cite{grill2007nano} and light \cite{grossmann2021surface} in combination with careful selection of molecular precursors has allowed for breakthrough work in the fabrication of carbon nanostructures and organic molecular networks \cite{grill2020covalent}. Partnered with the high-resolution characterisation SPM offers, sequences of on-surface reactions now provide molecular assembly options that are impossible in solution\cite{sweetman2020surface, zhong2021constructing, ren2023surface, sun2023surface}. Alongside this, the concept of single molecule engineering, to control all the elementary steps of a molecular chemical reaction via SPM manipulations, was first introduced in 2000 \cite{hla2000inducing}. Yet the potential of SPM for single molecule engineering has only emerged in recent years \cite{pavlicek2017generation,kaiser2019sp,zhong2021constructing,albrecht2022selectivity,ruan2024synthesis}. Despite these exciting results, it is clear that the technical challenges and time demands of manual manipulation approaches are not suitable for fabrication beyond a few molecules, and scaling these procedures beyond single manipulations and reactions to fabricate large molecular assemblies and engineer complex electronic states requires autonomous SPM operation\cite{gao2010scanning, sun2023surface}. 

In this paper, we establish a deep learning workflow to automate STM manipulations and optimize manipulation parameters to efficiently and selectively
break C--Br bonds in organobromides. Breaking these bonds is the first step of the Ullmann reaction \cite{clair2019controlling}, and an important intermediary step in OSS of complex molecules. This is then applied to Zn(II)-5,15-bis(4-bromo-2,6-dimethylphenyl)porphyrin (\ch{ZnBr2Me4DPP}) on Au(111) as a model system to study autonomous tip-induced reactions in STM. Meanwhile, Density Functional Theory (DFT)\cite{blum2009ab} calculations and Bayesian Optimization Structure Search (BOSS)\cite{todorovic2019bayesian} serve as an auxiliary tool to explore adsorption structures and reaction mechanisms in combination with SPM results and DRL models.

\section{Results and discussion}
\begin{figure}[!th]%
\centering
\includegraphics[width=\textwidth]{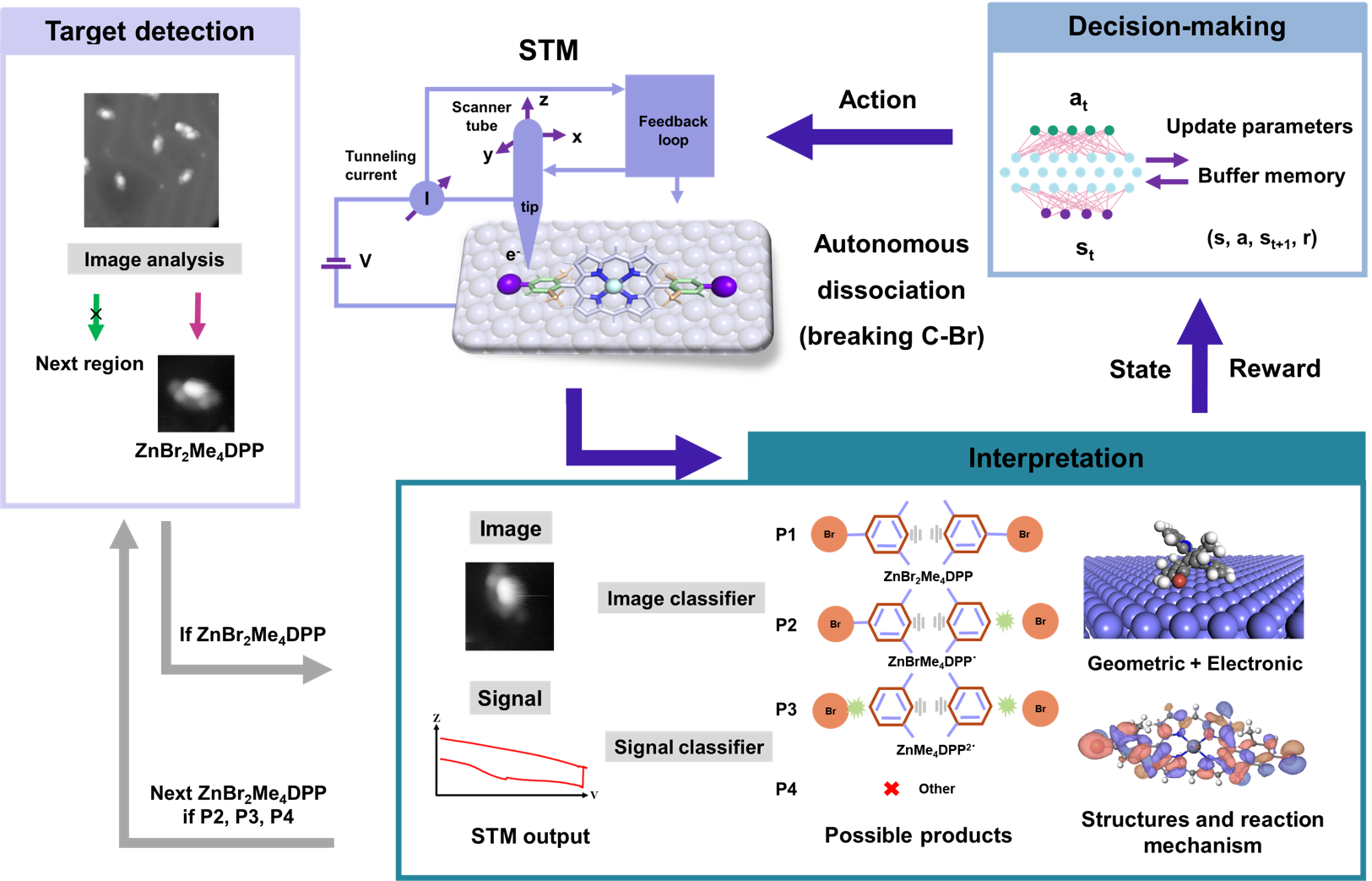}
\caption{\textbf{AutoOSS workflow}. AutoOSS consists of three key modules: target detection, decision-making and interpretation. The target detection module is responsible for detecting individual \ch{ZnBr2Me4DPP} candidate molecules from a larger scanning image by evaluating distances and areas of image contrast. The interpretation module aims at understanding the effect of manipulation parameters implemented on molecules through identifying products based on STM output (images and signals). The identification of products determines the next step. The decision-making module generates manipulation parameters. Here, we primarily employed two methods---a random generator and a DRL approach. The DRL approach searches for optimal STM manipulation parameters towards a goal using a reward system based on the state. Finally, a substantive number of 2D scanning images, reflecting various configurations of molecules on Au(111), collected during the whole process, can be used to analyze the geometric and electronic structures and potential reaction mechanisms with BOSS and DFT.}
\label{workflow}
\end{figure}

The overall architecture of our software infrastructure \textbf{AutoOSS} (\textbf{Auto}mated \textbf{O}n-\textbf{S}urface \textbf{S}ynthesis) consists of three components (Fig. \ref{workflow}): Target detection--search and identify targeted fragments based on STM images; Interpretation---Models to interpret the STM output during manipulation; Decision-making---DRL agent for selecting SPM parameters.

\newpage

\subsection {Target Detection}

\begin{figure}[!th]%
\centering
\includegraphics[width=\textwidth]
{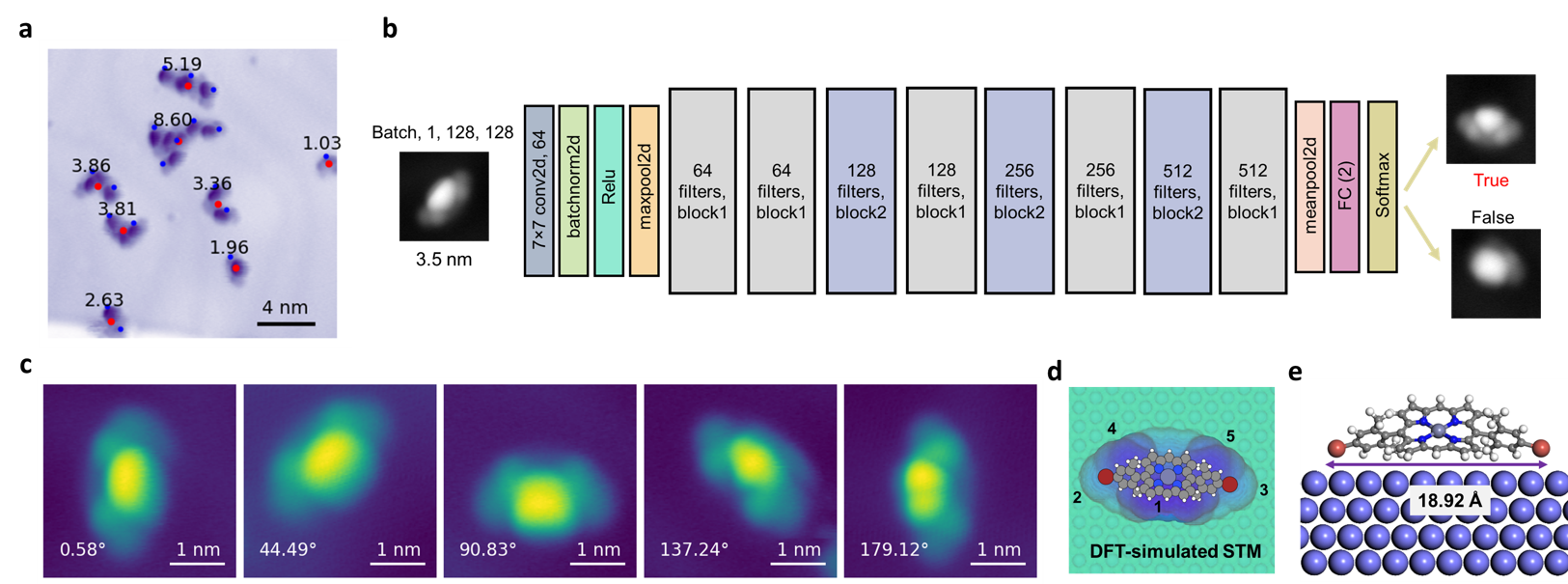}
\caption{\textbf{Search and identification of \ch{ZnBr2Me4DPP}}. (a) Detection of individual molecules from large images by the distance between molecules and the area of the contour. The blue points indicate the detected molecules in contours. Red points represent the centre point of contours, whose areas are marked by values. (b) Architecture of the neural network used to predict whether the image includes an individual \ch{ZnBr2Me4DPP}. (c) Example of targeted contrast patterns in STM images at different rotation angles. (d) 3D view of DFT simulated STM with superimposed molecular structure. (e) Side view of \ch{ZnBr2Me4DPP} adsorbed on Au(111).}
\label{detect}
\end{figure}

To efficiently detect promising candidate molecules to test C-Br bond dissociation, we acquired an STM image containing several molecules and molecular clusters (see Methods for details about the sample preparation). We then analyzed the distance between them (default: 2.5 nm, comparable to the size of molecules) and the area of the associated contrast patterns (default: 1.5 - 2.5 $\mathrm{nm^2}$) to exclude clusters or fragmented molecules in Fig. \ref{detect}a. However, many individual fragments share similar areas, especially the dissociated products resulting from the loss of one or two bromine atoms (Fig. S9a), which are hard to distinguish from one another. Therefore, we developed a neural network model to identify molecules more precisely based on magnified images focusing on the targeted patterns, where we zoomed in on a smaller scanning region of 3.5 $\times$ 3.5 nm -- still large enough to accommodate the target molecules measuring around 2.3 nm (Fig. \ref{detect}b). Furthermore, the complexity introduced by adsorbing a non-planar 3D structure onto a 2D surface, where the target molecule can undergo rotations and bind to the substrate at various sites and configurations, inevitably leads to diversity in observed STM contrasts. To understand the features of the molecule for target detection purposes, we correlated the observed STM contrast with multiple configurations aided by simulated STM images (see Methods and Fig. S5). Among these, we found that the most commonly seen contrast patterns in STM images (four lobes (2, 3, 4, 5) symmetrically around a larger lobe (1) in Fig. \ref{detect}c, d) match well with the three most stable adsorption structures (structures 1, 38, 73, 110, 115, 150 and 158 in Fig. S5), which have nearly isoenergetic computed energies ranging from -2.43 eV to -2.32 eV. 

To improve the ability of the models in identifying molecules, we defined the most frequently observed molecular features in the STM images as target objects while allowing minor deviations in tip condition and molecular rotation (Fig. \ref{detect}c). The 3D structure and corresponding STM images (Fig. \ref{detect}d, e) revealed that the central lobe 1 represents the upper periphery of the porphyrin ring, the two lobes of 2 and 3 at the ends are partly due to the presence of Br atoms, and the other two lobes of 4 and 5 originate mainly from the methyl fragments on the phenyl ring. Based on these characteristics, we manually constructed a labelled dataset of 1350 images, and an image classifier (Fig. \ref{detect}b) was trained to evaluate whether the scanning image includes an individual \ch{ZnBr2Me4DPP} -- the ultimate accuracy of the model was 98.5\% on the test dataset (more details in the method and Fig. S12).

\subsection{Interpretation}

\begin{figure}[!th]%
\centering
\includegraphics[width=\textwidth]
{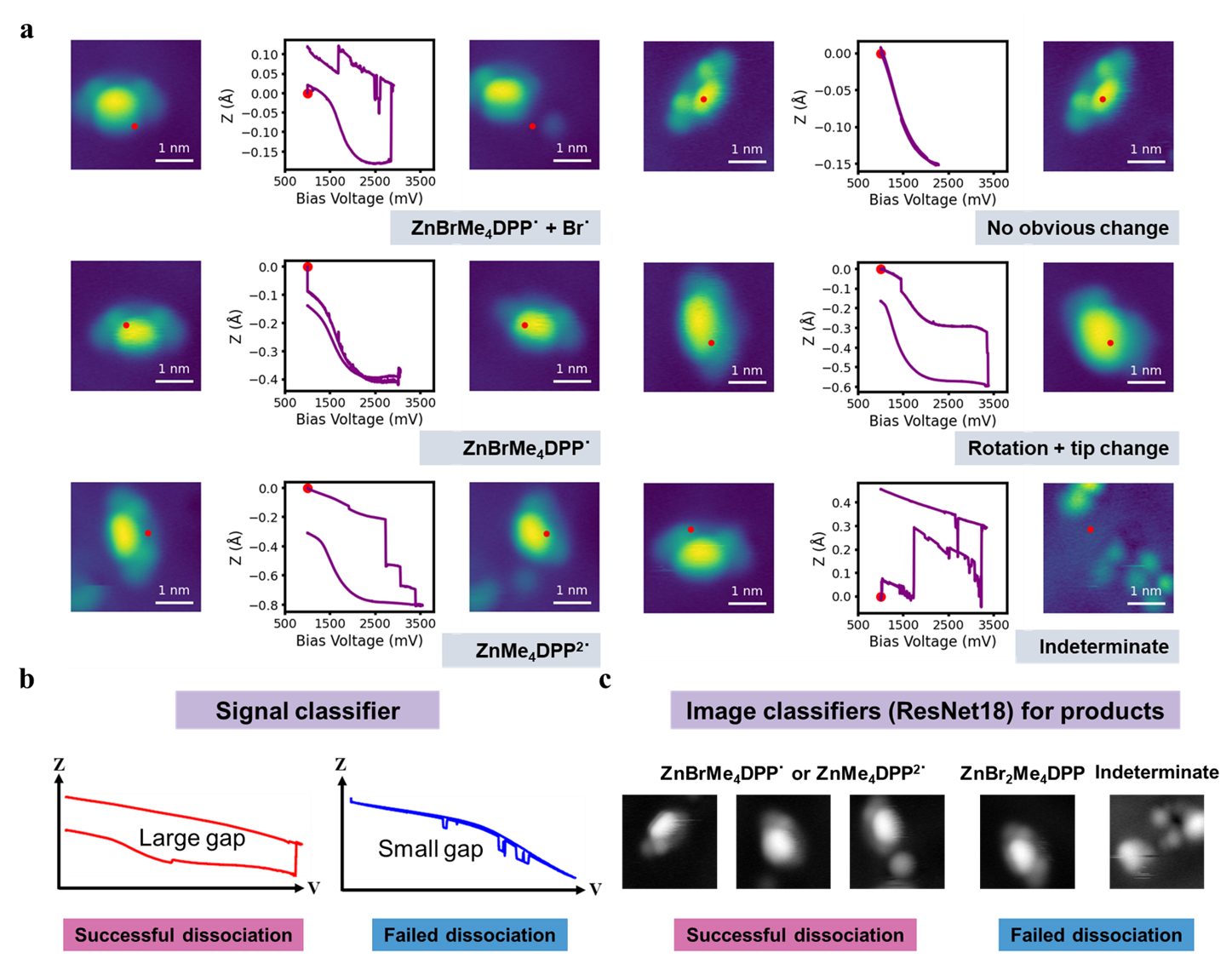}
\caption{\textbf{Interpretation of reaction.} (a) Possible states after dissociation. The images in the first and third columns refer to STM images before and after dissociation. Red points in images indicate the tip positions for dissociation. The second column represents the topography change during the implementation of parameters, where the red point marks the initial tip-sample distance. All values are relative to the initial tip-sample distance. (b) Illustration of signal classifier for evaluating whether the dissociation happens. (c) Three categories for evaluating products using image classifiers.}
\label{interpret}
\end{figure}

After finding and identifying the target \ch{ZnBr2Me4DPP} molecules, we are in a position to initiate the C-Br dissociation process by placing the STM tip on a specific site and applying a voltage bias (ramp pattern or pulse pattern, details shown in Fig. S16) and a current. Varying parameters among these four (tip$_x$, tip$_y$, V and I) may lead to various effects on the molecules, as shown by the representative selection in Fig. \ref{interpret}a. The dissociation of the C--Br bond(s) -- resulting in the corresponding dissociated molecules \ch{ZnBrMe4DPP^{.}} and \ch{ZnMe4DPP^{2.}} -- is the goal of the manipulation. However, as reflected in STM images, there are multiple possible outcomes of the manipulation process. For example, the contrast pattern of a Br atom (lobe 2 or lobe 3 in Fig. \ref{detect}d) may disappear or co-exist near the contrast patterns of \ch{ZnBrMe4DPP^{.}} or \ch{ZnMe4DPP^{2.}}. Besides, the appearance and contrast of these patterns may vary due to the possible changes in the STM tip apex during the manipulation process. In addition to changes in the chemical structure of the molecule, some manipulation parameters kept molecules intact simply resulting in its rotation or translation along the Au(111) surface as well as subtle shape or contrast changes due to different tip conditions. On the other hand, more extreme manipulation parameters can cause destructive damage to molecules and induce breaking of other bonds than C--Br, significant changes such as complete flips of the molecular configuration, large movements of molecules far away from the initial positions, and serious problems in tips like contamination, instabilities and multiple apices (see Fig. S8).

One of the major challenges for the automation of chemical reactions in SPM is to understand and recognize the consequences of applying manipulation parameters as outlined in the previous section. Due to the possibility of many complex outcomes, we opted to simply classify all of these into three categories: successful dissociation (Suc), intact molecule (Int) and indeterminate status (Ind), as shown in Fig. \ref{interpret}c.  This is used to determine if the manipulation action on a targeted molecule has to be continued (in the case of Int) or stopped (in the case of Suc and Ind), and whether the C--Br bond dissociation succeeds (in the case of Suc). For the Int category, the molecule may rotate but keep the typical characteristics of the target molecules, indicating that manipulation can continue. Meanwhile, Ind status and Suc status mean that the image pattern cannot be characterized as the targeted molecule any more, and the manipulation process is terminated. The difference between the two is that in the former we cannot establish if the C--Br bond has been dissociated, while in the latter it has clearly succeeded, either resulting in \ch{ZnBrMe4DPP^{.}} or \ch{ZnMe4DPP^{2.}}.

Aiming at automating this evaluation process of products, we analyzed over 5000 cases from the STM output (see Methods). The most straightforward way is to inspect the images after dissociation. Therefore, we trained classification models ($\mathrm{M_{Ind}}$ and $\mathrm{M_{Diss}}$) with experts labelling images to predict whether the products are Suc molecules (\ch{ZnBrMe4DPP^{.}} or \ch{ZnMe4DPP^{2.}}), \ch{ZnBr2Me4DPP} (Int molecules) or belonging to the Ind category (accuracy higher than 97 \%, more performance matrices and algorithmic details available in Fig S11-14 and Methods). 

Another obvious signal to consider as a classifier is the bias voltage (V)-topography (Z) curve -- it clearly exhibits different characteristics when resulting in different products during the dissociation (Fig. \ref{interpret}b). Generally, successful dissociation tends to be accompanied by a larger hysteresis in their V-Z curves. While a small hysteresis or even an overlapped curve emerges for manipulation parameters keeping molecules intact, especially those occurring at low voltage or current. Furthermore, we quantitatively estimated the three categories by analyzing the difference in topography between ramping up and down ($Diff_{topo}$) calculated by Eq. 5 in the section Signal classification. As shown in Fig. S11, there is some overlap in the distribution of the $Diff_{topo}$ for the three categories. However, the values among Int cases are usually smaller than 3.0 nm, and for Suc cases $Diff_{topo}$ values tend to be larger, even reaching 20 nm, whereas a broader range of $Diff_{topo}$ values (0 - 63 nm) is observed in Ind cases. While this offers useful insight into the dissociation process in some cases, we found that it was not a reliable classifier for DRL in general, as it was difficult to distinguish among different manipulation effects.

\subsection{Decision-making}

\subsubsection{Random action}

Developing models capable of interpreting STM manipulation outcomes is an essential precondition to finding the optimal parameters to reach the desired goal. Initially, we employed the most straightforward method --- random action --- to generate the four most relevant manipulation parameters for dissociation of the C--Br bonds: V, I, tip$_x$, tip$_y$. 

\begin{figure}[!th]%
\centering
\includegraphics[width=0.8\textwidth]
{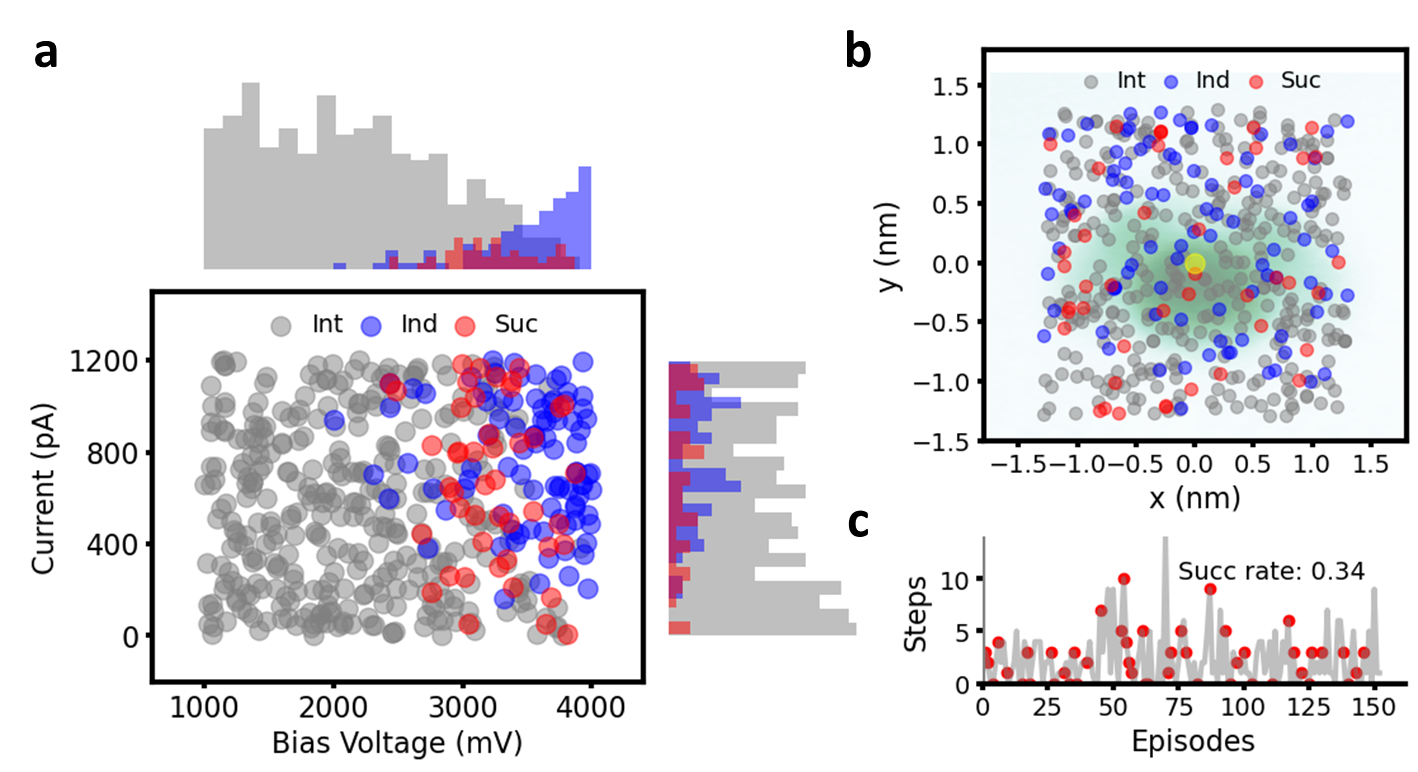}
\caption{\textbf{Performance of random action.} (a) Distribution of bias voltages and currents during 573 dissociation events. (b) Distribution of tip positions. A \ch{ZnBr2Me4DPP} molecule in green is superposed as a reference, where the yellow point represents the center point of the molecule, approximated as an ellipse. (c) The dissociation attempts for each episode before termination, where red points indicate successful dissociation.}
\label{perform_rand}
\end{figure}

We approximate the contrast pattern of a target molecule \ch{ZnBr2Me4DPP} as an ellipse, whose centre is defined as the reference tip position. Based on the size of the patterns in STM images (about 2.4 nm, Fig. S17), we limited the range of possible tip positions to within a radius of 1.3 nm from the reference position, sufficient to cover the whole pattern. In addition, the ranges of voltages and currents are set to 1200 - 4000 mV and 0 - 1200 pA based on domain knowledge. Fig. \ref{perform_rand} demonstrates the effects of 573 dissociation attempts on 150 molecules. Of these, 34\% of the molecules (Fig. \ref{perform_rand}c) were successfully dissociated into either \ch{ZnBrMe4DPP^{.}} or \ch{ZnMe4DPP^{2.}}, while the majority of molecules were categorized as Ind cases. The voltage and current distribution (Fig. \ref{perform_rand}a) revealed that successful dissociation reactions tend to occur at higher voltages (above 2400 mV), but are also accompanied with a high chance of unwanted reactions. However, the possibility of unwanted reactions could be reduced to some extent by using a lower current. We suspect that a higher current leads to multiple electrons being injected into the molecule, which excites multiple bonds, thus resulting in products which are difficult to analyze. Yet, the dependency of the applied current on the frequency of Ind cases is too noisy to make any clear conclusion in this regard.  Meanwhile, lower voltages sometimes result in rotation of the molecule, or no change at all. On the other hand, the dissociation reaction does not seem highly sensitive to the tip position, even when the tip is not directly on top of the molecule, it could still break the C--Br bond as desired. The d\textit{I}/d\textit{V} spectra detected at points 3 and 9 of the molecule (Fig. S7) indicate the characteristics of the Au(111) substrate, suggesting that the C--Br bond should be located between point 2 (or 8) and point 3 (or 9), whose distance referred to the centre point is less than 0.9 nm (Fig. S17). Moreover, we compared the result of the effect of random actions with the tip position constrained to be over molecules by reducing the radius from 1.3 nm to 0.6 nm in Fig. S16a -- the success rate slightly increased to 0.39. Meanwhile, the consequence of changing the voltage pattern from a pulse of 8 s to a ramp of 42 s demonstrated a comparable success rate of 0.40 in Fig. S16b (more details of the voltage patterns are illustrated in Methods, the pulse pattern is the default if not said otherwise). 

We further attempted to constrain the tip position near the C--Br bond based on Eq. 6, which ensures consistent positioning regardless of the rotational state of the contrast patterns in images, and also applied randomly generated bias voltages and currents to dissociate the molecule. The result for 164 molecules in Fig. S18 showed a similar trend in the distribution of voltage and current as with random tip position previously used. However, the success rate increases from 0.34 to 0.43, implying that specific tip positions could somewhat reduce the possibility of unwanted reactions of the molecules.

\subsubsection{Optimize action by DRL}

\begin{figure}[!th]%
\centering
\includegraphics[width=0.95\textwidth]
{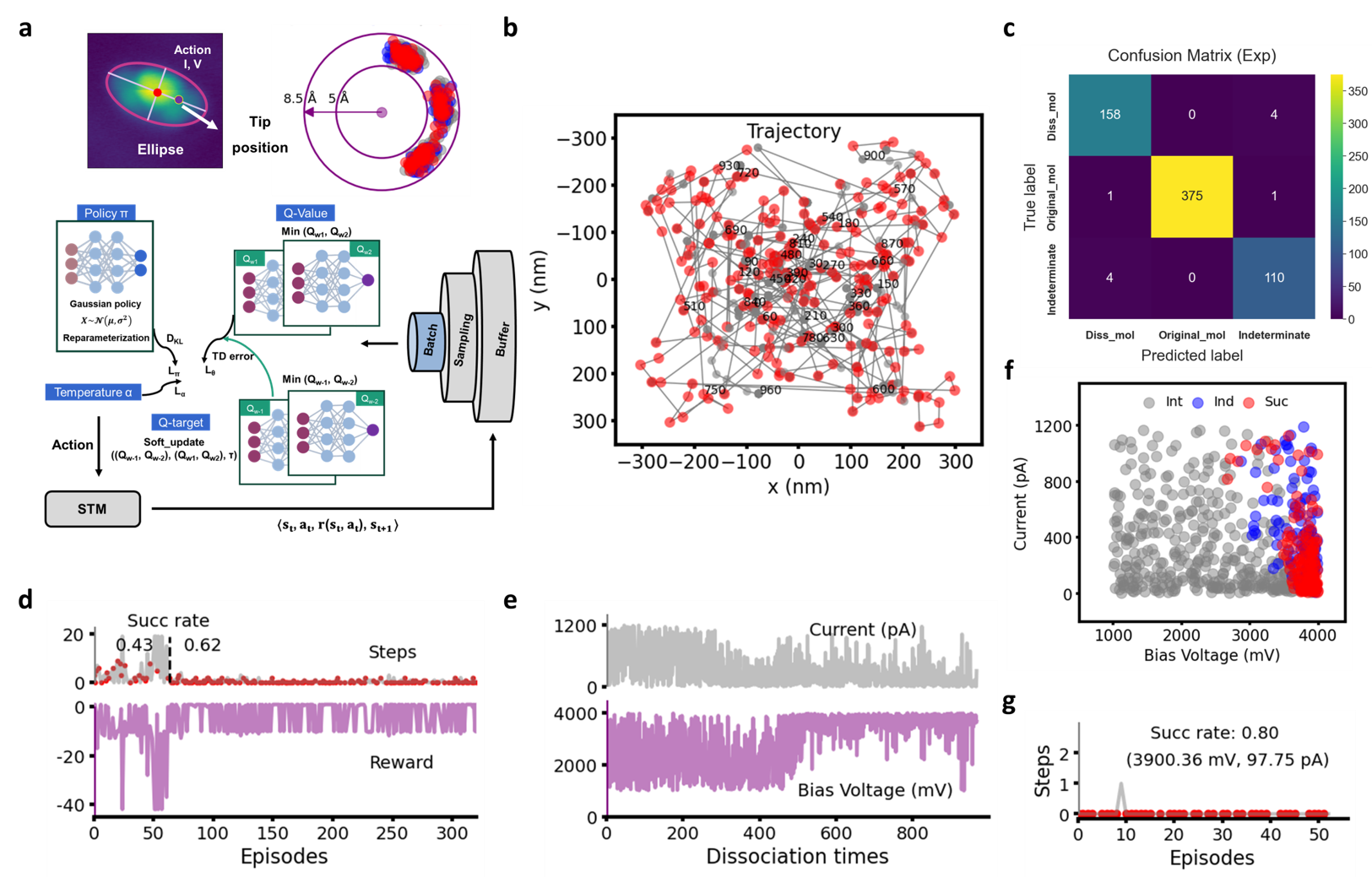}
\caption{\textbf{Performance of DRL model.} (a) Upper panel: illustration of the fixed tip position referred to the center point of a molecule, approximated as an ellipse (left) and the distribution of all tip positions referred to center points during DRL training process (right). down panel: architecture of DRL model based on the SAC algorithm. It consists of policy network, critic networks (Q-value function), temperature parameter and target Q-Networks and replay buffer. (b) Real STM trajectory while detecting targeted molecules \ch{ZnBr2Me4DPP} on the Au(111) sample and corresponding dissociation results. Red points indicate successful dissociation for the molecules, while gray points represent failed dissociation for the molecules after at most 20 attempts with varying various parameters. Here xy axes correspond to the STM measurement coordinates. (c) Performance of image classifier on unknown cases. (d) Evolution of of dissociate steps (top) and rewards (bottom) over episodes. (e) Evolution of bias voltages (top) and currents (bottom) over dissociation times. (f) Distribution of the pairs of voltage and current implemented on molecules for 968 dissociation times. (g) Repeatability test: dissociate 49 molecules at 3900.36 mV and 97.75 pA.}
\label{perform_rl}
\end{figure}

By definition, the random generator lacks the ability to optimize the dissociation parameters. Generally, this kind of decision-making problem can be formalized as a Markov decision process, where the manipulation parameters (action) depend solely on the current STM image (state). Therefore, we employed a DRL approach based on the Soft Actor-Critic (SAC) algorithm \cite{haarnoja2018soft} to optimize parameters for breaking the C--Br covalent bond, using a rational reward design based on interpreting the SPM scanning images during manipulations. To simplify the issue, we hypothesize states in DRL are the same with a 1D state space for all selected \ch{ZnBr2Me4DPP} molecules, regardless of tip condition and slight changes in the molecular conformations on the surface. The goal in our DRL models is to optimize bias voltage and current at the same specific tip position (Fig. \ref{perform_rl}a) under the reward system in Eq. 7. 

Fig. \ref{perform_rl}b displays the trajectory of 328 episodes with a total of 968 dissociation manipulations on \ch{ZnBr2Me4DPP} molecules. The $xy$ coordinates correspond to the real coordinates in the STM, reflecting the distribution of molecules in this region. The red points indicate molecules which underwent successful dissociation, while gray points indicate indeterminate cases. Note that the heterogeneity in success rate across the surface is a function of non-uniform distribution of molecules on the surface, interruptions in scanning for technical reasons and also the influence of regions used for tip conditioning, and it is difficult to make any inferences on the role of the surface itself. The confusion matrix in Fig. \ref{perform_rl}c further confirms the high accuracy of image classifier models on the unknown dataset. Fig. \ref{perform_rl}d illustrates that the model starts to converge after 60 episodes, with a success rate before 60 episodes of 0.43, consistent with that in the tests using random voltage and current. After 60 episodes, the success rate increases to 0.62 and the dissociation steps per episode are fewer than 3 in most cases, also with a higher occurrence of larger accumulated rewards. The fluctuation of rewards between 1 and -10 could be attributed to the high proximity between parameters leading to successful and indeterminate dissociation. Due to differences in molecular conformations on Au(111) and tip condition, parameters that lead to successful C-Br dissociation for one molecule may result in an indeterminate dissociation for another. This is confirmed by repeatedly testing these successful parameters to dissociate molecules (Fig. S19), where the success rate is just 0.42, comparable to that in random dissociation. Meanwhile, the voltage converged to higher values (more than 3500 mV) whereas the current narrows to lower values (less than 400 pA), despite some fluctuations, as shown in Fig. \ref{perform_rl}e, f. Such narrowing of the range of voltage and current guided by the reward, decreases the dissociation steps per episode and increases the success rate. Furthermore, the analysis for the distribution of these parameters in Fig. S20 implies that the sets of parameters with voltage higher than 3800 mV and current less than 200 pA offer a higher possibility to successfully dissociate molecules and reduce indeterminate cases. 

To explore whether we can further increase the success rate, we randomly selected a set of parameters with a lower current value (97.75 pA) and higher voltage value (3900.36 mV) from those associated with a high likelihood of successful dissociation in DRL training, and repeatedly applied these parameters to 49 molecules,  as shown in Fig. \ref{perform_rl}g, obtaining an increase of the success rate up to 0.8. This demonstrates the feasibility of our model applied on long-term, selective, efficient operations for autonomous on-surface synthesis in STM. Furthermore, the orbital energies of the highly localized C--Br $\sigma$* states of the adsorbed \ch{ZnBr2Me4DPP} molecule with respect to the Fermi level at 3.7 eV (Fig. S10e), comparable to the voltage bias applied to promote successful dissociation, suggest that selective bond dissociation is probably achieved by tunneling electrons into the corresponding antibonding states, consistent with previous literature \cite{lesnard2007dehydrogenation,pavlicek2017generation,kawai2020three}.  

\section{Conclusions}

To summarize, we have demonstrated the capability of a deep learning model to identify reactants and products based on STM outputs, enabling a DRL agent to evaluate various manipulation parameters. Furthermore, the establishment of a deep reinforcement learning approach allows the agent to optimize these parameters. These advancements address key challenges in STM automation and molecular synthesis. Ultimately, the integration of target detection module, interpretation module and decision-making module into the AutoOSS workflow achieved the automation of tip-induced C--Br bond breaking from \ch{ZnBr2Me4DPP} in STM. AutoOSS enables long-term, selective and efficient operations without human intervention. Moreover, the extensive dataset accumulated from experiments, combined with big-data analysis, DFT calculations and BOSS, offers the opportunity to uncover hidden physical information, explore 3D molecular conformations and investigate reaction mechanisms despite the limits of resolution in STM images. 

AutoOSS paves the way for automating manipulations in on-surface synthesis, thus pioneering a new paradigm in single molecular engineering. Moving forward, we anticipate the possibility of extending AutoOSS to a diverse array of molecules and applications pertinent to complex chemical reactions, encompassing various chemical bonds, molecules, tip, and manipulation types. For similar reaction processes, moving to different molecule and substrate combinations requires only retraining the model on appropriate classifiers of reaction success. Furthermore, there is the potential to enhance the model's selectivity and precision by using a refined tip, optimized bias voltage pattern or incorporating AFM signals into the workflow to provide atom-level resolution scanning images.

\section{Methods}
\subsection{Experimental preparation and STM microscopy}\label{method1}

\begin{figure}[!th]%
\centering
\includegraphics[width=0.5\textwidth]
{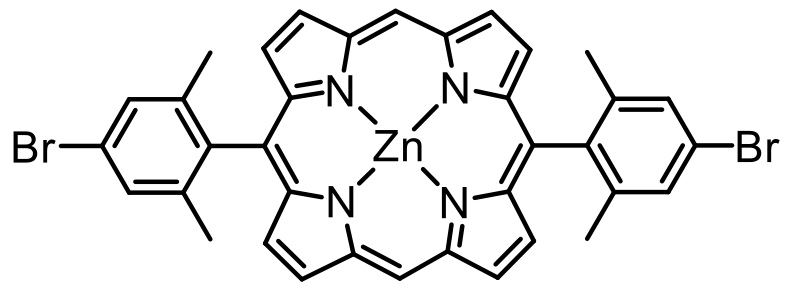}
\caption{Chemical structure of \ch{ZnBr2Me4DPP}.}
\label{chem_struct}
\end{figure}

\ch{ZnBr2Me4DPP} molecules (chemical structure shown in Fig. \ref{chem_struct}) were synthesized via the precursors 5,15-bis(4-bromo-2,6-dimethylphenyl)porphyrin (\ch{H2Br2Me4DPP}) from 2,6-dimethyl-4-bromobenzaldehyde, as shown in Fig. S22. Characterizations associated with \ch{ZnBr2Me4DPP} molecules and precursors \ch{H2Br2Me4DPP} (Fig. S23-S30), including \textsuperscript{1}H and \textsuperscript{13}C nuclear magnetic resonance (\textsuperscript{1}H-NMR and \textsuperscript{13}C-NMR) spectroscopy, mass spectrometry (MS) and ultraviolet-visible (UV/vis) spectroscopy, were implemented. Then, \ch{ZnBr2Me4DPP} molecules were evaporated from a Knudsen cell heated to 230 \textcelsius{} onto a Au(111) sample kept below 7 K temperature. The STM scaned and dissociations were performed in constant current mode on a Createc LT-STM system with a gold-coated PtIr tip. The STM images recorded at different scales from 100 nm to 3.5 nm are shown in Fig. S1. Contrast-adjusted STM images in Fig. S3 are available to show the adsorption sites of individual molecules on various regions either in the herringbone of the Au(111) or on the Au(111) surface. Ultimately, we chose 20 $\times$ 20 nm to detect promising targeted molecules and 3.5 $\times$ 3.5 nm to make further identification by neural networks and dissociation manipulations. The scanning speed and the number of pixels for all images are 1000 \AA/s and 128, resulting in around 42 s per image.

\subsection{Spiral path planning}\label{method2}

The approach area was about 700 $\times$ 700 nm, where four 100 $\times$ 100 nm squares near the boundary were set aside to form tips. Therefore, the manipulation region usually corresponded to the XY coordinates in STM from -300 nm to 300 nm, where the centre point of the region of 20 $\times$ 20 nm for detecting target candidates was updated by the shortest distance $d_{dist}$ away from the reference point beyond forbidden area. It was formulated as: $d_{dist}=d_{Eucli} + \alpha * d_{Manha}$, where $d_{Eucli}$ and $d_{Manha}$ indicate Euclidean distance $d_{Eucli}= \sqrt{(x-x_{ref})^2+(y-y_{ref})^2}$ and Manhattan distance $d_{Manha}=|x-x_{ref}|+|y-y_{ref}|$, respectively, and the coefficient of $\alpha$ is set as 1. 

\subsection{Detect target candidates}\label{method3}

We firstly converted raw scanning images from STM to greyscale images and then made further analyses to detect target candidates using two methods. One method is to limit the distance between image contrast patterns, where binary images with the threshold pixel values of 50 (pixels less than 50 were set to 0) and 150 (pixels greater than 150 were set to 255) were obtained to find the individual molecules through a thresholding distance (default: 2.5 nm) between points to get rid of dimer, trimer or clusters. Another method is to limit the area of patterns, for which we detected contours by the Otsu algorithm\cite{cheriet1998recursive, gong1998fast} with a clear outline (another two algorithms---global thresholding and Otsu thresholding after Gaussian filtering were compared by corresponding areas in Fig. S2). Based on statistical analysis of candidates, we restricted the area of patterns within 1.5-3.0 $\mathrm{nm^2}$ to further exclude some individual fragments. 

\subsection{BOSS}\label{method4}

We employed the Bayesian Optimization Structure Search (BOSS) method \cite{todorovic2019bayesian} to reduce the number of DFT evaluations needed to map out the configurational phase space. Data points were initialized with a quasi-random Sobol sequence and the GP-Lower Confidence Bound acquisition function with increasing exploration (elcb) were used on all runs. The kernels for rotation and $xy$-translation were standard periodic kernel (stdp), while the $z$-coordinate used radial basis functions (rbf). The surface symmetry was exploited to multiply the acquired data points by applying symmetry operations to the adsorbate at high-symmetry sites, where the Au(111) surface has three rotationally equivalent sites in addition to two translationally equivalent ones. Initially, a conformational search was conducted on the isolated gas-phase \ch{ZnBr2Me4DPP} molecule, with the search variables being full rotation of the phenyl moieties and their methyl substituents (6D search). The surrogate model was constructed on 407 DFT data points. The search resulted in one single main conformer in terms of phenyl rotation, as shown in Fig. S4, which was subsequently employed as the molecular building block in the adsorption structure search. The same  structure was used as the building block for \ch{ZnBrMe4DPP^{.}} and \ch{ZnMe4DPP^{2.}}, since loss of the terminal Br atoms does not result in significant rearrangement of the rest of the molecule following DFT relaxation. The adsorption structure search was done by constructing a surrogate model of the DFT (PBE+vdW$^{surf}$) PES for the translational and rotational degrees of freedom (6D search), and subsequently relaxing the lowest energy surrogate model local minima with DFT, thus accounting for any changes in the structures of the isolated rigid molecular building blocks enabled by the surface interaction. The molecular adsorbate building blocks for the search were the lowest energy \ch{ZnBr2Me4DPP}, \ch{ZnBrMe4DPP^{.}}, \ch{ZnMe4DPP^{2.}} species as described above, which were combined with the relaxed 11$\times$12$\times$4 Au(111) substrate building block. The surrogate models for the adsorption structures were constructed out of 262, 94 and 108 data entries for \ch{ZnBr2Me4DPP}, \ch{ZnBrMe4DPP^{.}}, and \ch{ZnMe4DPP^{2.}}, respectively. The global minimum predictions were oscillating between the symmetrically equivalent rotational configurations ($\pm$60$^{\circ}$).

\subsection{DFT calculations}\label{method5}

All DFT computations were performed using FHI-aims \cite{blum2009ab}. For the initial conformational search using BOSS, we employed the B3LYP functional \cite{becke1988density, lee1988development} with light defaults and first tier basis functions. Subsequently, the resulting global minimum conformers, substrate, and all adsorption structures were relaxed to a forces less than 0.01 eV/\AA$^2$ using the Perdew-Burke-Ernzerhof (PBE) functional augmented with the van der Waals dispersion correction, including collective screening effects of the substrate electrons (vdW$^{surf}$), fully denoted (PBE+vdW$^{surf}$) \cite{tkatchenko2009accurate, ruiz2012density}. This choice of functional for both isolated and adsorbed molecules was motivated by the properties of an adsorbed configuration being of interest, for which this functional has been demonstrated accurate in comparison with experiments \cite{liu2015quantitative}. The same functional was also used during BOSS data acquisition iterations for the adsorption structures. The Brillouin zone was sampled using a 1$\times$1$\times$1 Monkhorst-Pack grid, and the slab was constructed using four layers of 11$\times$12 gold atoms as the Au(111) surface, of which the two lowest layers were kept fixed during all computations. The length of the box in the $z$-direction was in total 60 \AA, ensuring sufficient vacuum space. This relatively large slab size was chosen to avoid interactions with adjacent adsorbates. Spin polarization was used for the dissociated \ch{ZnBrMe4DPP^{.}} and \ch{ZnMe4DPP^{2.}} species.

The STM images were simulated using FHI-aims with the Tershoff-Hamann approximation as implemented therein \cite{tersoff1985theory}. The simulation bias was kept at 1.0 V for all images, which were created with VESTA \cite{momma2011vesta} using an isovalue between 10$^{-10}$ and 10$^{-12}$ a.u. to match experimental STM images.

The calculations for the C--Br dissociation model reactions were performed using the climbing image nudged elastic band and growing strings methods \cite{jonsson1998nudged, peters2004growing} as implemented in https://gitlab.com/cest-group/aimsChain-py3. The pathways of both bond cleavage reactions were modeled using 12 images in total, where the growing strings force threshold was 0.5 eV/\AA$^2$, while the climbing image threshold was 0.05 eV/\AA$^2$. The initial structure for the reaction was the global minimum adsorption configuration as determined by BOSS, while the dissociated final structures of each step in the reaction were determined by moving the Br atom 5 \AA$ $ away from the rest of the porphyrin, and relaxing with DFT to the force threshold as the initial image.  

The adsorption energy is formulated as: $E_{ads} = E_{mol+sub}-E_{mol}-E_{sub}$, where \textit{mol+sub} denotes molecule on the substrate, \textit{mol} the isolated molecule and \textit{sub} the isolated substrate.

\subsection{Image classification}\label{method6}

All image classifiers were developed based on the ResNet18 model \cite{he2016deep} (the architecture of the neural network shown in Fig. \ref{detect}b and Fig. S6), taking STM images with a size of 3.5 $\times$ 3.5 nm and the pixel numbers of 128 as input. To ensure the intact pattern of fragment in the image, we adjust the scanning region based on the centre of the pattern in STM images and scan again if the centre point is beyond the threshold region. The default criterion for the centre in the pattern is less than 0.438 nm along both $xy$ axes, referred to as the centre point of the scanning region.  

The image classifiers consist of three binary models ($\mathrm{M_{Target}}$, $\mathrm{M_{Ind}}$, $\mathrm{M_{Diss}}$) and one multi-class model ($\mathrm{M_{Triple}}$) with the numbers of corresponding datasets shown in Tab. 1, intended for detecting reactants and distinguishing products. Due to the complexity in products, caused by variable tip conditions, various conformations as well as subtle differences for dissociated molecules and pristine molecules, we trained another two binary models ($\mathrm{M_{Ind}}$ and $\mathrm{M_{Diss}}$) for more elaborate distinctions to supplement $\mathrm{M_{Triple}}$. In brief, these models were designed to distinguish \ch{ZnBr2Me4DPP}  and non \ch{ZnBr2Me4DPP} ($\mathrm{M_{Target}}$), to distinguish indeterminate and non indeterminate ($\mathrm{M_{Ind}}$) and to distinguish intact molecules (\ch{ZnBr2Me4DPP}) and dissociated molecules (\ch{ZnBrMe4DPP^{.}} or \ch{ZnMe4DPP^{2.}}) ($\mathrm{M_{Diss}}$). 

The Adam optimizer \cite{kingma2014adam} with cross entropy loss function and StepLR were used to optimize parameters in models. In addition, Bayesian optimization was introduced to optimize learning rate based on the converged loss values under corresponding learning rate values. Eventually, the optimal learning rates in Adam optimizer are $1.11\times e^{-5}$, $4.22\times e^{-5}$, $5.84\times e^{-5}$, 0.0001 for $\mathrm{M_{Target}}$, $\mathrm{M_{Ind}}$, $\mathrm{M_{Diss}}$, $\mathrm{M_{Triple}}$, respectively. All models perform decently with the accuracy more than 94\%  and the under the curve (AUC) higher than 98\% (more performance metrics are shown in Fig. S12-15 and Tab. S1-4). A confusion matrix divides the classification results into 4 categories through comparing the true values and predicted values: True Position (TP, both real values and predicted values are 1), True Negative (TN, both real values and predicted values are 0), False Positive (FP, real values are 0, but predicted values are 1), False Negative (FN, real values are 1, but predicted values are 0).

\begin{equation}
    \begin{aligned}[b]
    Accuracy & = \frac{TP+TN}{TP+FP+TN+FN}     
    \end{aligned}
\end{equation}

\begin{equation}
    \begin{aligned}[b]
    Precision & = \frac{TP}{TP+FP}     
    \end{aligned}
\end{equation}

\begin{equation}
    \begin{aligned}[b]
    Recall & = \frac{TP}{TP+FN}     
    \end{aligned}
\end{equation}

\begin{equation}
    \begin{aligned}[b]
    F1 & = \frac{2*Precision*Recall}{Precsion+Recall}     
    \end{aligned}
\end{equation}

\begin{table}[htb]
\captionsetup{justification=centering} 
\caption{Dataset for four images classifiers \label{tab_class}}
\begin{tabular}{cccccc}
\hline
\multicolumn{3}{c}{\textbf{Target or not}}                                  & \multicolumn{3}{c}{\textbf{Indeterminate or not}}     
\\ 
\textbf{class}                   & \textbf{train} & \textbf{test}             & \textbf{class}                   & \textbf{train}       & \textbf{test} \\ \hline
True mol                      & 273                  & 39                   & Indeterminate                         & 1186            & 270           \\ 
Non true                      & 1116                 & 159                  & Non indeterminate                     & 2764           & 607           \\ \hline
\multicolumn{3}{c}{\textbf{Dissociation or not}}                            & \multicolumn{3}{c}{\textbf{Products}}                          \\ 
\textbf{class}                & \textbf{train}       & \textbf{test}        & \textbf{class}                & \textbf{train} & \textbf{test} \\ \hline
\multicolumn{1}{l}{Original mol} & 2023                  & 438                   & \multicolumn{1}{l}{Original mol} & 2023            & 438            \\ 
\multicolumn{1}{l}{Diss mol}  & 741                  & 169                   & \multicolumn{1}{l}{Diss mol}  & 741            & 169            \\ 
\multicolumn{1}{l}{}          & \multicolumn{1}{l}{} & \multicolumn{1}{l}{} & \multicolumn{1}{l}{Indeterminate}     & 1186            & 270            \\ \hline                      

\end{tabular}
\end{table}

\subsection{Signal classification}\label{method7}

We tested two types of voltage bias patterns---a ramp of 42s and a pulse of 8s, with similar processes, as shown in Fig. S16. Times were divided into 1024 steps. The initial voltages are 1 V: for a ramp pattern, the voltage starts to increase to the specific voltages from point 20 until point 512, symmetrically, then decrease to 1 V at point 1004; while for a pulse pattern, the voltage directly jumps to the specific voltage at point 20, which is maintained until point 1004, then back to 1 V. We analyzed signal changes during the dissociation by the difference of topography, formulated as follows:

\begin{equation}
Diff_{topo} = |\sum_{i=1}^{512} V_{topo}-\sum_{i=513}^{1024} V_{topo}| 
\end{equation}

\noindent where $V_{topo}$ indicates the value of topography at a point during the process of voltage variation along 1024 points.

\subsection{Specific tip position}\label{method8}

\begin{equation}
tip_{x}, tip_{x} = \gamma \times H \times \sin{\alpha} + \beta_{1}, -\gamma \times H \times \cos{\alpha} + \beta_{2}
\end{equation}

\noindent where H and $\alpha$ is the height and the angle of an ellipse evaluated by the fitEllipse function in OpenCV \cite{bradski2008learning}, $\gamma$ is a coefficent (default: 0.3), $\beta_{1}$ and $\beta_{2}$ are random noise ranging from -0.1 \AA $ $ to 0.1 \AA.

\subsection{Reward design}\label{method9}

The assessments from image classifiers are applied to  evaluate the reward and make further decisions. The reward is defined as:

\begin{equation}
  r_{t}(s_{t}, s_{t+1}) =
    \begin{cases}
      1-(factor)*t & \text{\text{\ch{ZnBrMe4DPP^{.}}} or \text{\ch{ZnMe4DPP^{2.}}}}\\
       
      -0.2-(factor)*t & \text{\text{\ch{ZnBr2Me4DPP}}}\\

      -10-(factor)*t & \text{\text{Indeterminate}}\\

    \end{cases}       
\end{equation}

\noindent where \textit{t} indicates the dissociate times in an episode, and \textit{factor} is a coefficient with a default value of 0.2.

\subsection{Success rate}\label{method10}

The success rate in a test is defined as:

\begin{equation}
  succ \ rate = \frac{N_{Suc}}{N_{Succ}+N_{Ind}}
\end{equation}

\noindent In this equation, $N_{Suc}$ and $N_{Ind}$ represent, respectively, the number of successful dissociation and the number of the indeterminate cases for all episodes in one test. The value is used to evaluate the ability of the model optimizing parameters to avoid indeterminate cases. Dissociation steps in an episode can be used to assess how fast the model can find successful dissociation parameters. Therefore, the unchanged dissociation is not necessarily considered here.  

\subsection{Tip conditioning}\label{method11}

The tip may suffer bluntness, contamination, instability, damage or multiple tips during the scanning process. Correspondingly, different parameters related to voltage, indentation depth and time may be needed to identify the condition for a sharp and stable tip. To maintain a good tip, the workflow monitors the tip condition and reforms when needed. Experience in this task demonstrates that a random approach heights ranging from 2 nm to 5.5 nm and constant voltage of 1 V usually can achieve a decent tip. As the criteria for successful tip conditioning, we search for candidate molecules in a 20 $\times$ 20 nm image -- if this fails many times (default: 4), our algorithm tries deeper immersion with 10 nm. This strategy allows a long-term operations in the whole workflow ranging from detection to dissociation. Tip conditioning is activated only when detecting molecules. Once the targeted molecule is found, it consecutively tests different dissociation parameters until it terminates, without the interruption from conditioning tip. On one hand, the movement during the tip conditioning may lead to the shift of coordinates in STM and the tip status after forming may be complex and still effectively bad, which may damage targeted molecules. On the other hand, if the molecules are damaged by a dissociation manipulation rather than tip itself, we treat it as a failed manipulation, and classify it as a indeterminate case. Furthermore, empirically, gentle manipulations during dissociation in our task sometimes even make tip better in obtaining high-quality scanning images. Therefore, tip conditioning during dissociation attempts is not necessary.

To avoid moving the tip long distances, four square regions with the length of 100 nm near the edge of approach area were set, among which the tip moves towards the closest one for conditioning. To reduce the time on scanning the tip conditioning region, we just condition the tip at a random point in the square in practice. As an option, the algorithm of detecting point from the blank region to avoid molecules (Fig. S21) is available.

\subsection{Soft Actor-Critic}\label{method12}

The SAC approach consists of policy sampling module for mapping a state to an action, two state value q network for evaluating the state value, and one state-action value q network. The maximum entropy RL in the model maximizes the cumulative rewards and also pursues the diversity of policy through introducing the entropy term:

\begin{equation}
    \begin{aligned}[b]
    V(s_{t}) & = E_{a_{t} \thicksim \pi} [Q(s_{t}, a_{t}) - \alpha log\pi(a_{t}|s_{t})]  
             & = E_{a_{t} \thicksim \pi} [Q(s_{t}, a_{t})] + H( \pi(\cdot|s_{t}))   \label{eqt:sac}     
    \end{aligned}
\end{equation}

\begin{equation}
  \pi_{new} = arg\, \underset{{\pi}'}{min}\,  D_{KL}({\pi }'\left ({\cdot |s}  \right ), \frac{exp\left ( \frac{1}{a}Q^{\pi_{old}}\left ( s,\cdot  \right ) \right )}{Z^{\pi _{old}}\left (s,\cdot   \right )}) \label{eqt:policy} 
\end{equation}

\noindent 1. Value network: The loss function of value:
\begin{equation} 
    \begin{aligned}[b]
      L_Q(\omega) & = E_{s_{t}, a_{t}, r_{t}, s_{t+1}  \thicksim R \,}[\frac{1}{2}(Q_{w}(s_{t}, a_{t})-(r_{t}+\gamma V_{w^{-}}(s_{t+1})))^2] \\
                  & = E_{s_{t}, a_{t}, r_{t}, s_{t+1}  \thicksim R \,}[\frac{1}{2}(Q_{w}(s_{t}, a_{t})-(r_{t}+\gamma (E_{a_{t+1} \thicksim \pi_{\theta}} [Q(s_{t+1}, a_{t+1})] \\
                  & + H( \pi(\cdot|s_{t+1}))))^2] \\
                  & = E_{s_{t}, a_{t}, r_{t}, s_{t+1}  \thicksim R, \, a_{t+1} \thicksim \pi_{\theta}(\cdot | s_{t+1}) }[\frac{1}{2}(Q_{w}(s_{t}, a_{t})-(r_{t}+\gamma (min Q_{w^{-}}(s_{t+1}, a_{t+1}) \\
                  & - \alpha log \pi(a_{t+1}|s_{t+1})))^2]          \label{eqt:value} 
    \end{aligned}
\end{equation}

\noindent 2. Policy network: The action was determined by policy network, which generated the mean and std of Gaussian distribution by 1 linear layer  (hidden\_dim: 512) and then sampled action from the Gaussian distribution. The loss function of policy was set based on the Kullback-Leibler (KL) Divergence\cite{Kullback1951}:


\begin{equation}
  L_\pi(\theta) = E_{s_{t}  \thicksim R, \, a_{t} \thicksim \pi_{\theta}}[\alpha log \pi(a_{t}|s_{t}) - min Q_{w}(s_{t}, a_{t})]
\end{equation}

\noindent 3. Entropy regularization: To maximize the entropy, the corresponding loss function was set as follows:

\begin{equation}
  L(\alpha) =  E_{s_{t}  \thicksim R, \, a_{t} \thicksim \pi_{\theta}}[-\alpha log \pi(a_{t}|s_{t}) - \alpha H_{0}]
\end{equation}

\noindent where $\alpha$ is the temperature parameter.

In addition, the advanced sampling technique, Hindsight Experience Replay (HER)\cite{andrychowicz2017hindsight},  was used to improve the data efficiency. The optimal hyperparameters found in testing are learning rate lr of 0.0003, discount factor $\gamma$ of 0.99 and target smoothing pf $\tau$ of 0.1.

\begin{acknowledgement}

We thank the valuable discussion and assistance from Roman Fasel, Bruno Schuler, Hongxiang Xu, Shuning Cai, Zhenrong Zhao, Xueyong Jia, Zhengmao Li, Mohammad Amini and Büşra Gamze Arslan. The authors acknowledge fundings from the Academy of Finland (project no. 318995, 320555, 346824, 347319). A.S.F. was supported by the World Premier International Research centre Initiative (WPI), MEXT, Japan. This research was part of the Finnish centre for Artificial Intelligence FCAI. The authors acknowledge the computational resources provided by the Aalto Science-IT project and CSC, Helsinki. T.T. ad G.B. acknowledge financial support from the Spanish MCIN/AEI/10.13039/501100011033 and European Union NextGenerationEU/ PRTR (PID2020-116490GB-I00, TED2021-131255B-C43), MICIU /AEI /10.13039/501100011033 / FEDER, UE (PID2023-151167NB-I00), the Comunidad de Madrid and the Spanish State through the Recovery, Transformation and Resilience Plan [“Materiales Disruptivos Bidimensionales (2D)” (MAD2D-CM) (UAM1)-MRR Materiales Avanzados], and the European Union through the Next Generation EU funds. IMDEA Nanociencia is appreciative of support from the “Severo Ochoa” Programme for Centers of Excellence in R\&D (CEX2020-001039-S).

\end{acknowledgement}

\begin{suppinfo}

The Supporting Information is available free of charge. 

Detection of individual \ch{ZnBr2Me4DPP}: STM images at different scales, Measuring the area of contrast patterns, Individual molecules on Au(111), Conformational analysis of \ch{ZnBr2Me4DPP}, Various configurations of \ch{ZnBr2Me4DPP} on Au(111), ResNet18 Block, dI/dV spectra and maps of \ch{ZnBr2Me4DPP} on Au(111);

Interpretation: Example of indeterminate cases, Properties of contrast patterns in STM images, Reaction energies and possible dissociation mechanisms, Signal classifier, Performance metrics of $\mathrm{M_{Target}}$, Performance metrics of $\mathrm{M_{Triple}}$, Performance metrics of $\mathrm{M_{Ind}}$, Performance metrics of $\mathrm{M_{Diss}}$;

Decision-making: Random action at different bias patterns, Contrast pattern measurement, Random action with fixed tip position, Random selection from successful dissociation parameters, Dissociation parameters from the DRL model, Tip conditioning region;

Synthesis and characterization: Materials and Methods, Synthesis of precursors \ch{H2Br2Me4DPP} and \ch{ZnBr2Me4DPP}, as well as their corresponding \textsuperscript{1}H-NMR, \textsuperscript{13}C-NMR, MS and UV/vis.

\end{suppinfo}

\section*{Data and Software availability}
A video demonstrating AutoOSS's ability to autonomously and selectively control the reaction, all training dataset and parameters in machine learning models and input and output of BOSS and DFT calculations can be obtained on the Zenodo repository at \url{https://doi.org/10.5281/zenodo.13761822}.  The source codes and examples are available on the GitHub repository at
\url{https://github.com/SINGROUP/AutoOSS}.


\bibliography{achemso-demo}

\end{document}


\newpage

\textbf{List of abbreviations}

AcOEt = Ethyl acetate

AFM = Atomic Force Microscopy

AUC = Area under the curve 

BOSS = Bayesian Optimization Structure Search

CREST = Conformer-Rotamer Ensemble
Sampling Tool

DFT = Density Functional Theory 

DCM = Dichloromethane

DCTB = trans-2-[3-(4-tert-butylphenyl)-2-methyl-2-propenylidene]malononitrile

DDQ = 2,3-Dichloro-5,6-dicyano-1,4-benzoquinone

DRL = Deep reinforcement learning 

HER = Hindsight Experience Replay

MALDI-TOF = Matrix-assisted laser desorption/ionization-time of flight

MS = Mass spectrometry

NMR = Nuclear magnetic resonance

OSS = On-surface synthesis

Rf = Retention factor

RL = Reinforcement learning 

SAC = Soft Actor-Critic

SPM = Scanning Probe Microscopy

STM = Scanning Tunnelling Microscopy

THF = Tetrahydrofurane

UV/vis = Ultraviolet-visible

\ch{H2Br2Me4DPP} = 5,15-bis(4-bromo-2,6-dimethylphenyl)porphyrin

\ch{ZnBr2Me4DPP} = Zn(II)-5,15-bis(4-bromo-2,6-dimethylphenyl)porphyrin

\newpage

\section{Detection of individual \ch{ZnBr2Me4DPP}}

\subsection{STM images at different scales}

\begin{figure}[!th]%
\centering
\includegraphics[width=\textwidth]
{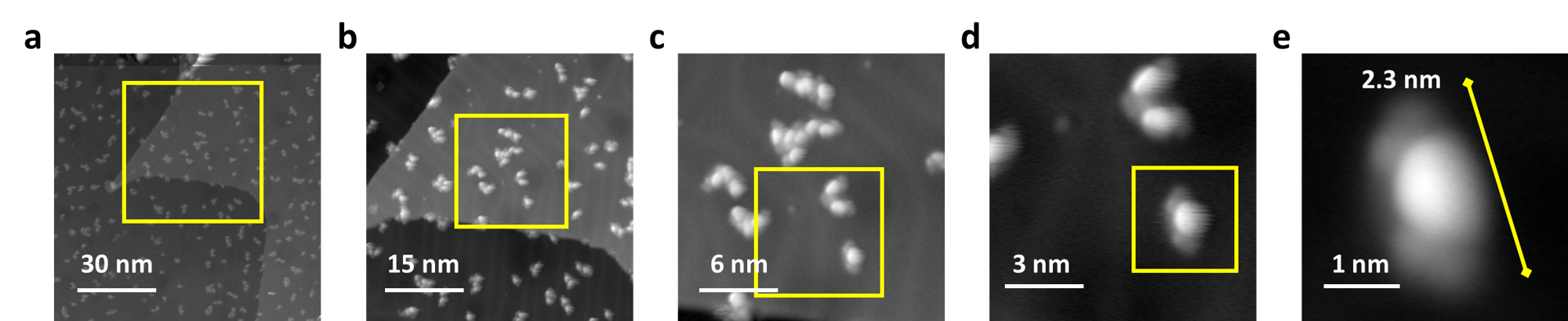}
\caption{\textbf{STM images at difference scales.} (a) 100 nm $\times$ 100 nm, (b) 50 nm $\times$ 50 nm,  (c) 20 nm  $\times$ 20 nm,  (d) 10 nm $\times$ 10 nm,  (e) 3.5 nm $\times$ 3.5 nm.}s
\label{different size}
\end{figure}

\subsection{Measuring the area of contrast patterns}

\begin{figure}[!th]%
\centering
\includegraphics[width=0.6\textwidth]
{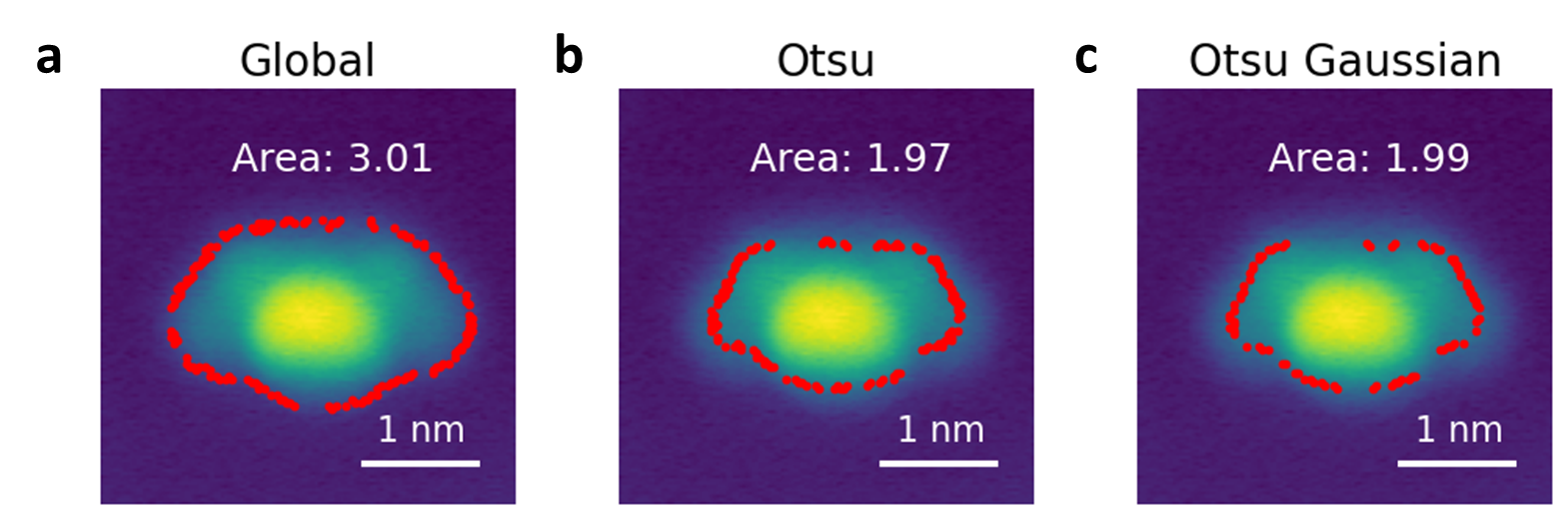}
\caption{\textbf{Comparison of three image thresholding methods.} (a) Global, (b) Otsu and (c) Otsu Gaussian for detecting the contrast pattern in an STM image.}
\label{measure area}
\end{figure}

\newpage

\subsection{Individual molecules on Au(111)}

\begin{figure}[!th]%
\centering
\includegraphics[width=0.9\textwidth]
{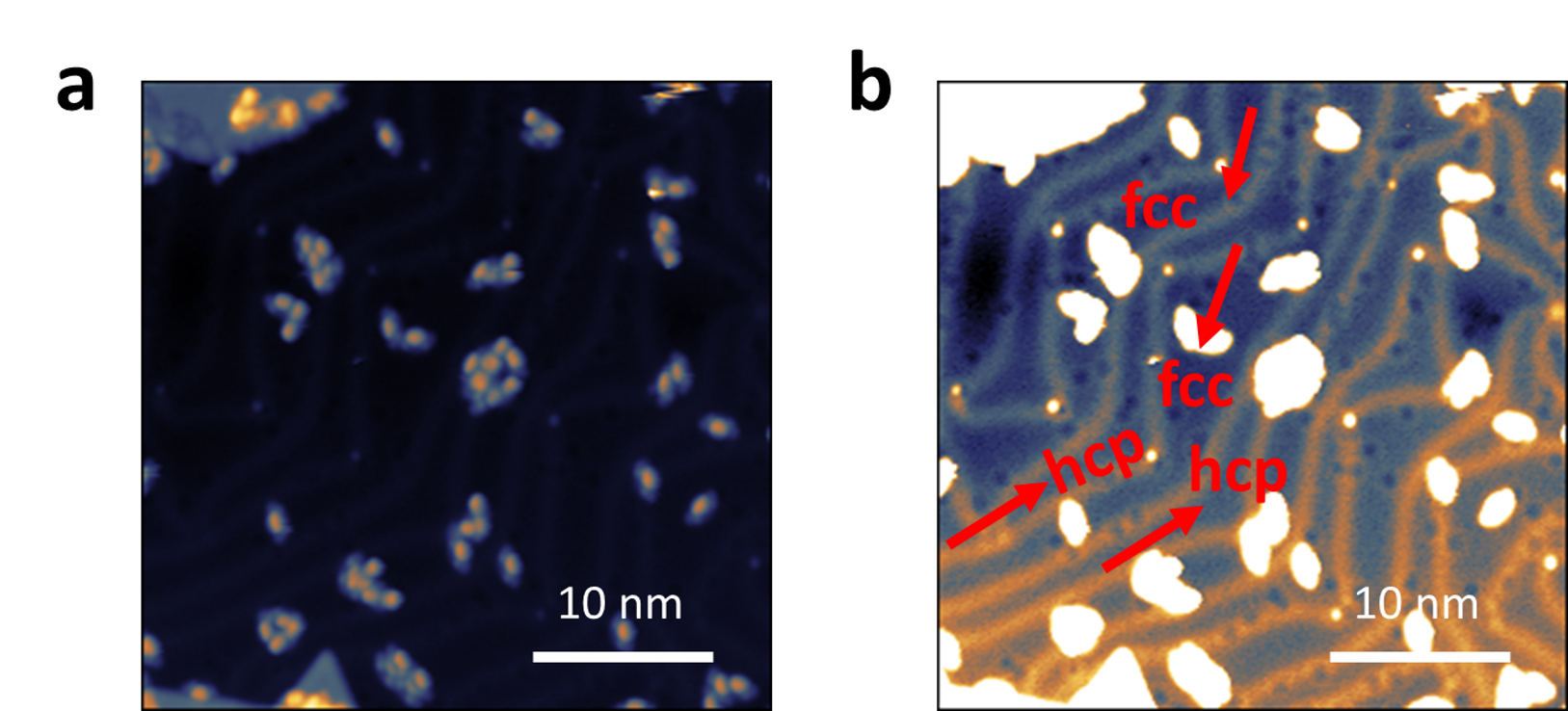}
\caption{(a, b) STM image of target molecules adsorbed at the hexagonal close-packed (hcp), face-centered cubic (fcc) and elbow regions of the reconstructed Au(111) surface \cite{edmondson2022molecular}. In (b), the herringbones are visible by adjusting the image contrast.}
\label{herringbone}
\end{figure}

\newpage

\subsection{Conformational analysis of \ch{ZnBr2Me4DPP}}

\begin{figure}[!th]%
\centering
\includegraphics[width=0.95\textwidth]
{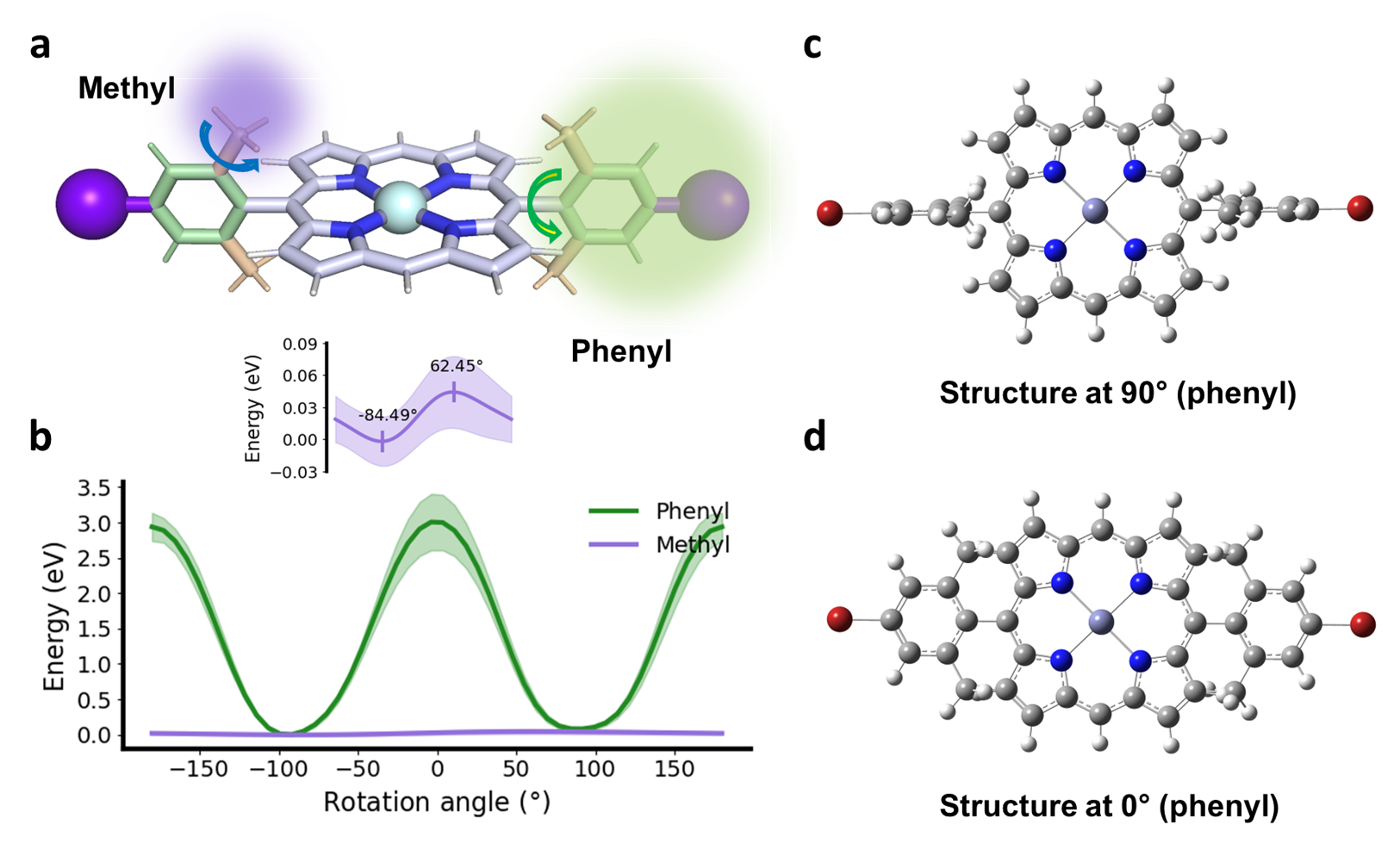}
\caption{\textbf{Conformational analysis of \ch{ZnBr2Me4DPP} using BOSS.} (a) Schematic illustration of rotatable C(porphyrin)-C(phenyl) bond (purple shadow) and C(phenyl)-C(methyl) bond (green shadow). (b) Corresponding energy diagrams by scanning rotation angles. The inset magnifies the energy curve for the rotation around the C(phenyl)-C(methyl) bond.  (c) Structure with the phenyl group at $90^{\circ}$ angles around the C(porphyrin)-C(phenyl) bonds. (d) Structure with the phenyl group at $0^{\circ}$ (Methyl groups in phenyls bump into the porphyrin ring).}
\label{rotation}
\end{figure}

The most important conformational change for the \ch{ZnBr2Me4DPP} molecule is due to rotation of the \ch{C6H2(CH3)2Br} (phenyl) groups relative to the central porphyrin. It should be noted that free rotation of the \ch{CH3} (methyl) groups around the C-C bond that connect them to the phenyl is possible, but this rotation does not change the overall conformations of the porphyrin enough to warrant a separate analysis of the corresponding rotamers on the surface, as these minute structural changes are likely superseded by those induced by adsorbate-substrate interactions. The main conformer of \ch{ZnBr2Me4DPP} has the phenyl moieties orthogonal to the central porphyrin backbone, with free rotation being hindered by the methyl groups on the former (rotation barrier around 3.0 eV corresponding to clash between atoms, see Fig. \ref{rotation}b). Futhermore, the backbone of the molecule has some limited flexibility, but since changes in this structure is most accurately described by collective motion of the molecule, it is difficult to assign simple variables required by BOSS, such as bond stretch, bending or torsion to this motion. Therefore, we also used the Conformer-Rotamer Ensemble Sampling Tool (CREST)\cite{Pracht2020}, which employs metadynamics including collective variables to check for alternative gas phase conformers. However, this analysis also yielded one conformer only, in addition to a large number (around 300) of similar rotamers, corresponding to the aforementioned collective movement of the porphyrin backbone, the phenyl side groups, and the methyl groups.

\newpage

\subsection{Various configurations of \ch{ZnBr2Me4DPP} on Au(111)}

\begin{figure}[!th]%
\centering
\includegraphics[width=0.95\textwidth]
{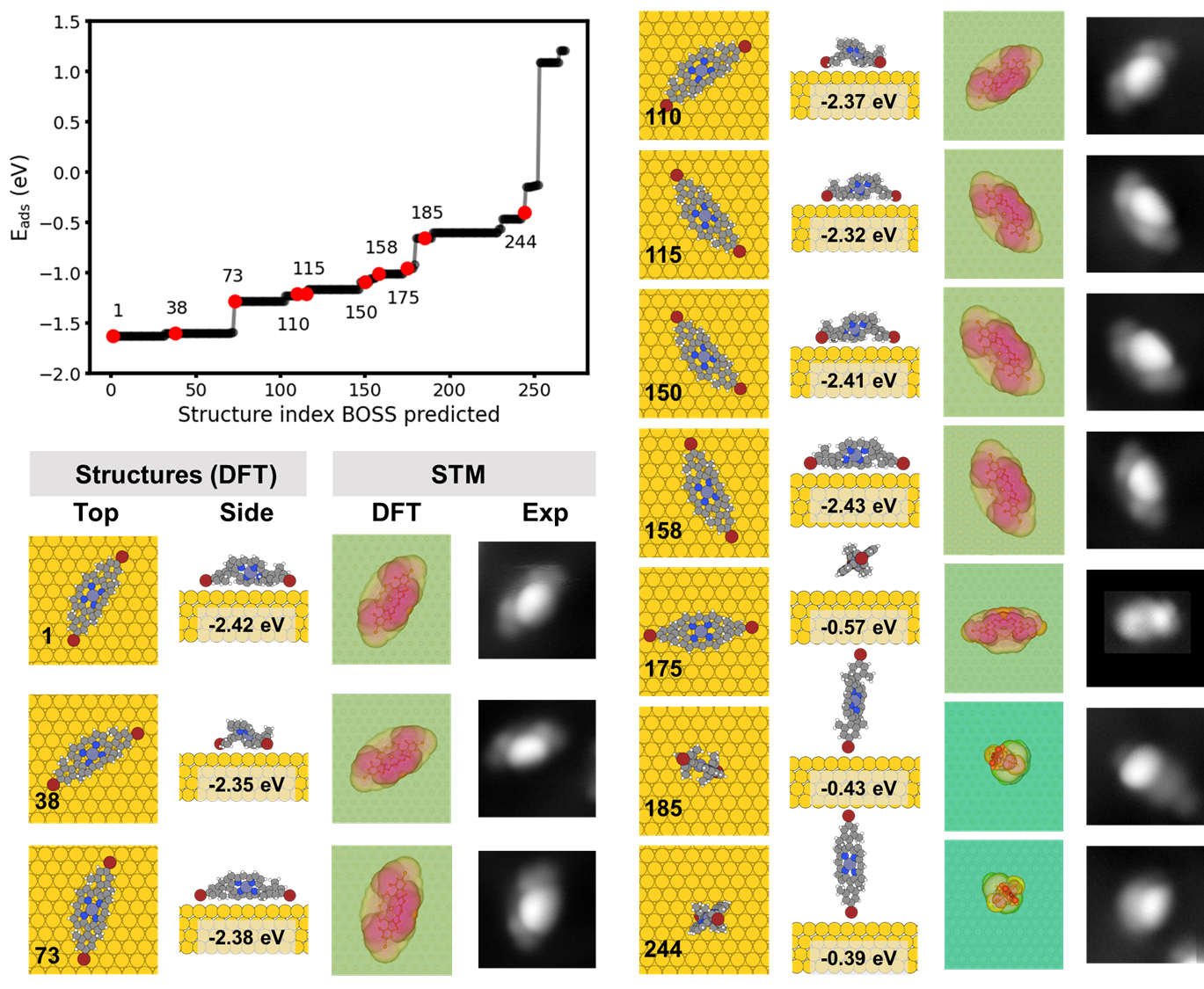}
\caption{\textbf{Various configurations of \ch{ZnBr2Me4DPP} adsorbed on Au(111) searched by BOSS.} The energy diagram represents the searched structures and corresponding adsorption energies obtained from BOSS. Some stable configurations with possible corresponding STM scanning images are shown. From left to right, the columns represent 3D structures optimized by DFT calculations from top view and from side view, DFT-simulated STM, experimental scanning images (setpoint: 1$\times10^{-11}$ A, bias voltage: 1 V). The values on the side structures indicate the adsorption energies obtained by DFT calculations.}
\label{boss}
\end{figure}

\clearpage
\newpage

\subsection{ResNet18 Block}

\begin{figure}[!th]%
\centering
\includegraphics[width=0.8\textwidth]
{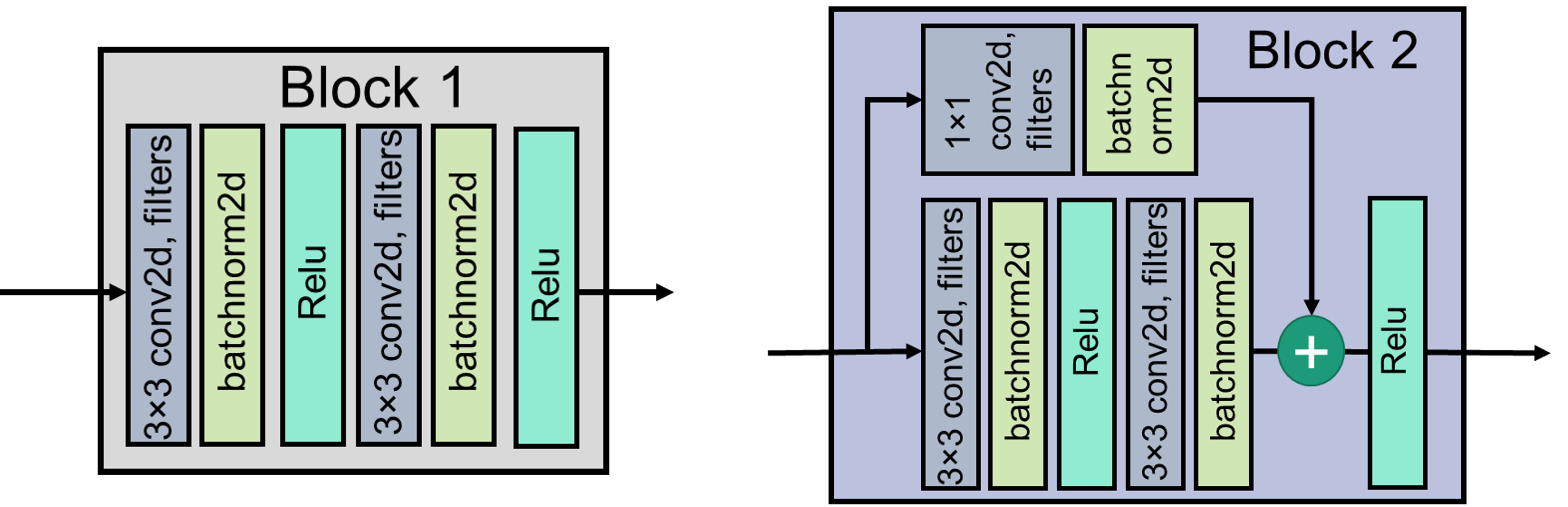}
\caption{\textbf{Architectures of Block 1 and Block 2 in ResNet18.} Block 1 is stacked by two sets of convolution layer, normalization layer and Relu layer, which is a regular ResNet block. Block2 enhances a regular ResNet block by incorporating a 1$\times$1 convolution with a normalization.}
\label{resnet block}
\end{figure}

\newpage

\subsection{d\textit{I}/d\textit{V} spectra and maps of \ch{ZnBr2Me4DPP} on Au(111)}

\begin{figure}[!th]%
\centering
\includegraphics[width=0.8\textwidth]
{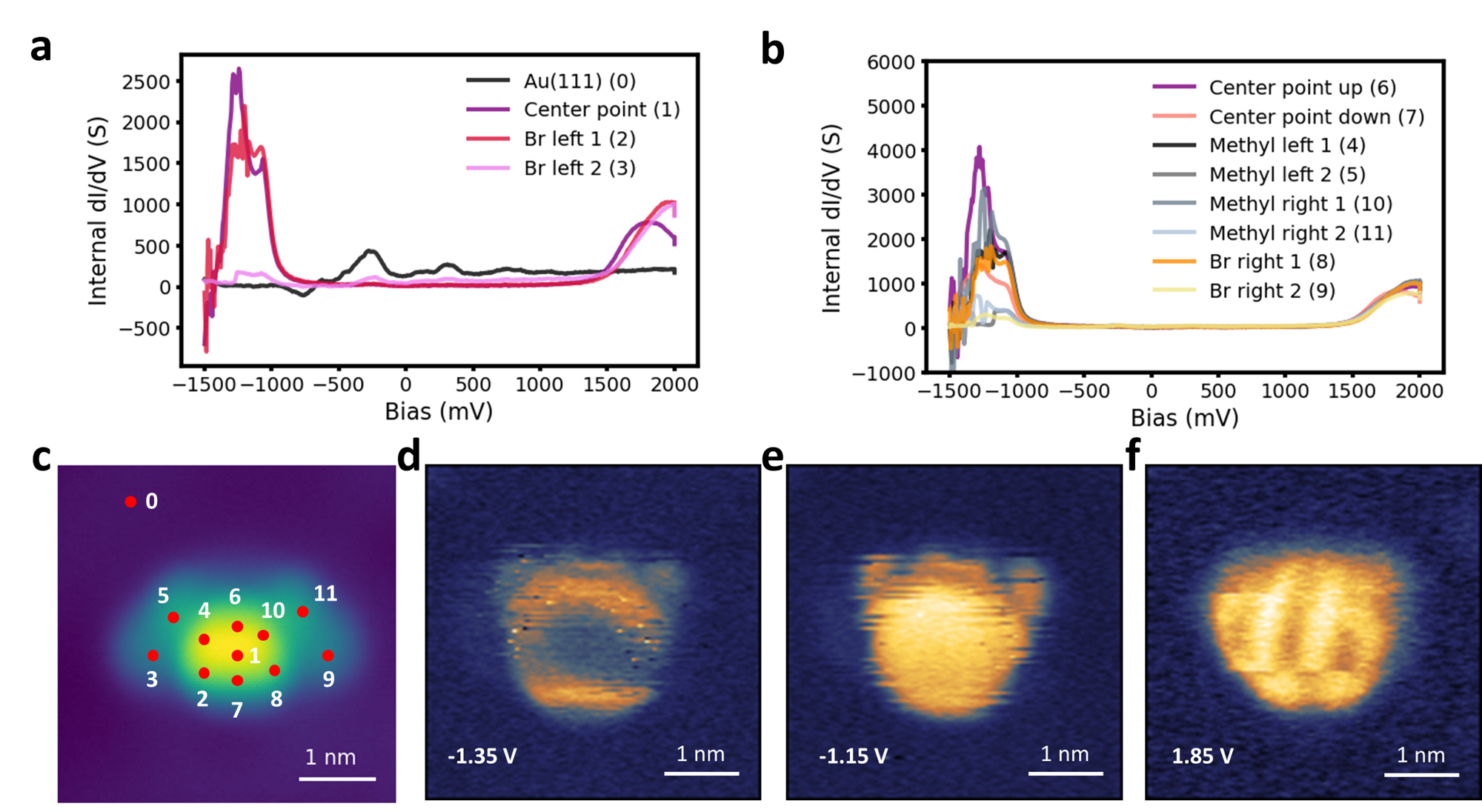}
\caption{(a,b) Experimental d\textit{I}/d\textit{V} spectra at points 0 - 11 on the pattern corresponding to a \ch{Br2Me4DPP} molecule in (c) an STM image (STS feedback loop at 5$\times10^{-10}$ A and 2 V for all cases). (d, e,f) dI/dV maps recorded at -1.35 V (HOMO-1), -1.15 V (HOMO) and 1.85 V (LUMO), the setpoint is 5$\times10^{-11}$ A, constant current mode.}
\label{didv}
\end{figure}

\newpage

\section{Interpretation}

\subsection{Example of indeterminate cases}

\begin{figure}[!th]%
\centering
\includegraphics[width=0.95\textwidth]
{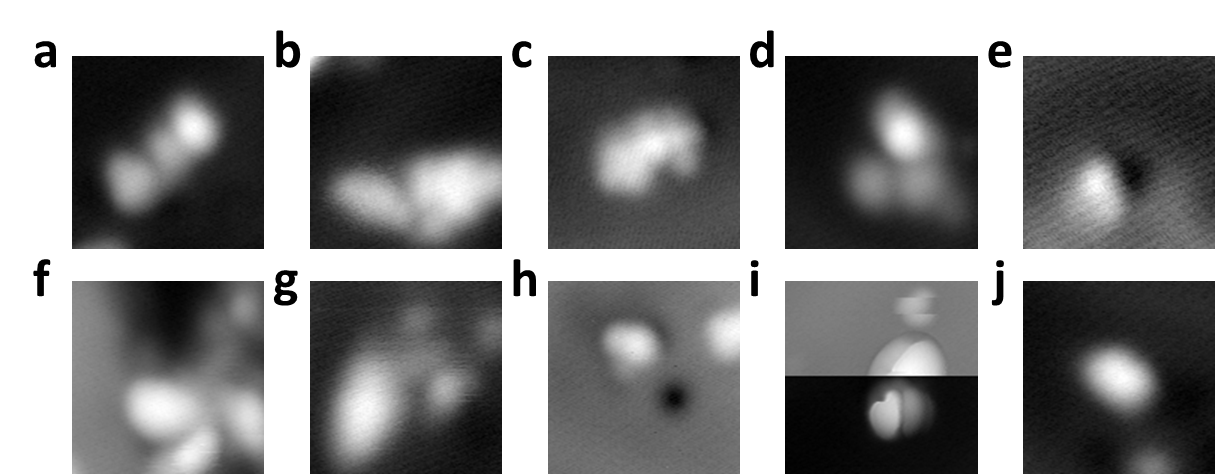}
\caption{Example of indeterminate cases in STM images.}
\label{ind_cases}
\end{figure}

\subsection{Properties of contrast patterns in STM images}

\begin{figure}[!th]%
\centering
\includegraphics[width=\textwidth]
{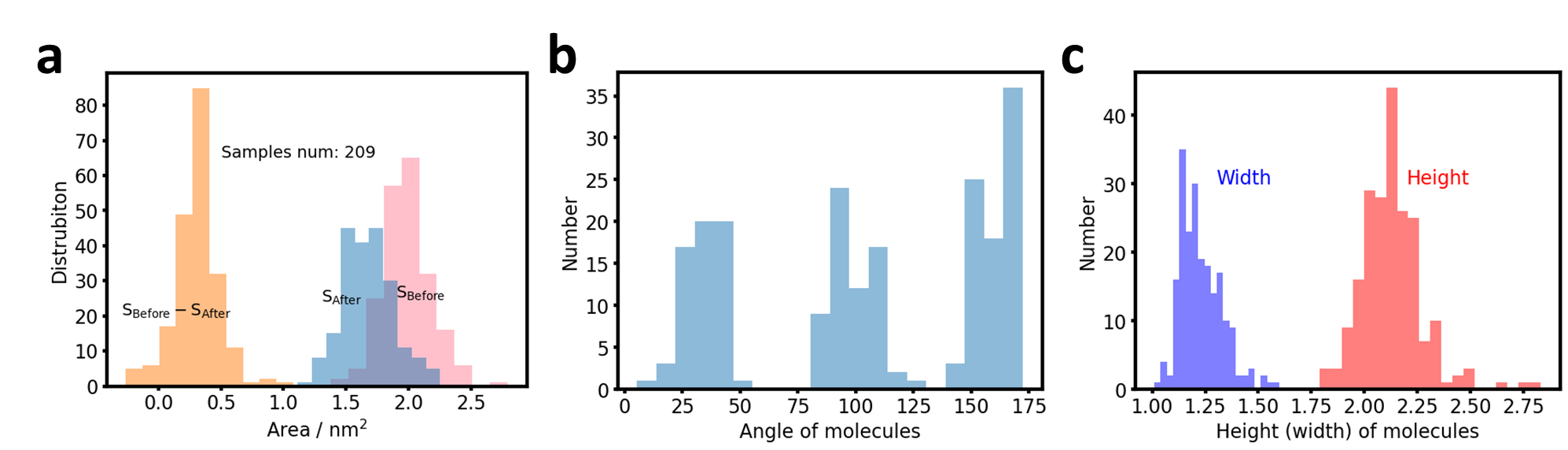}
\caption{\textbf{Properties of contrast patterns in STM images.} (a) Distribution of areas of patterns of pristine \ch{ZnBr2Me4DPP} and their corresponding products (\ch{ZnBrMe4DPP^{.}} or \ch{ZnMe4DPP^{2.}}) after implementing manipulations. (b) Distributions of angles, (c) heights and widths for the approximated ellipses from  patterns of reactants \ch{ZnBr2Me4DPP}. Data obtained by analysing 209 pairs of images among successful cases.}
\label{property of fragments}
\end{figure}

\newpage

\subsection{Reaction energies and possible dissociation mechanisms}

\begin{figure}[!th]%
\centering
\includegraphics[width=\textwidth]
{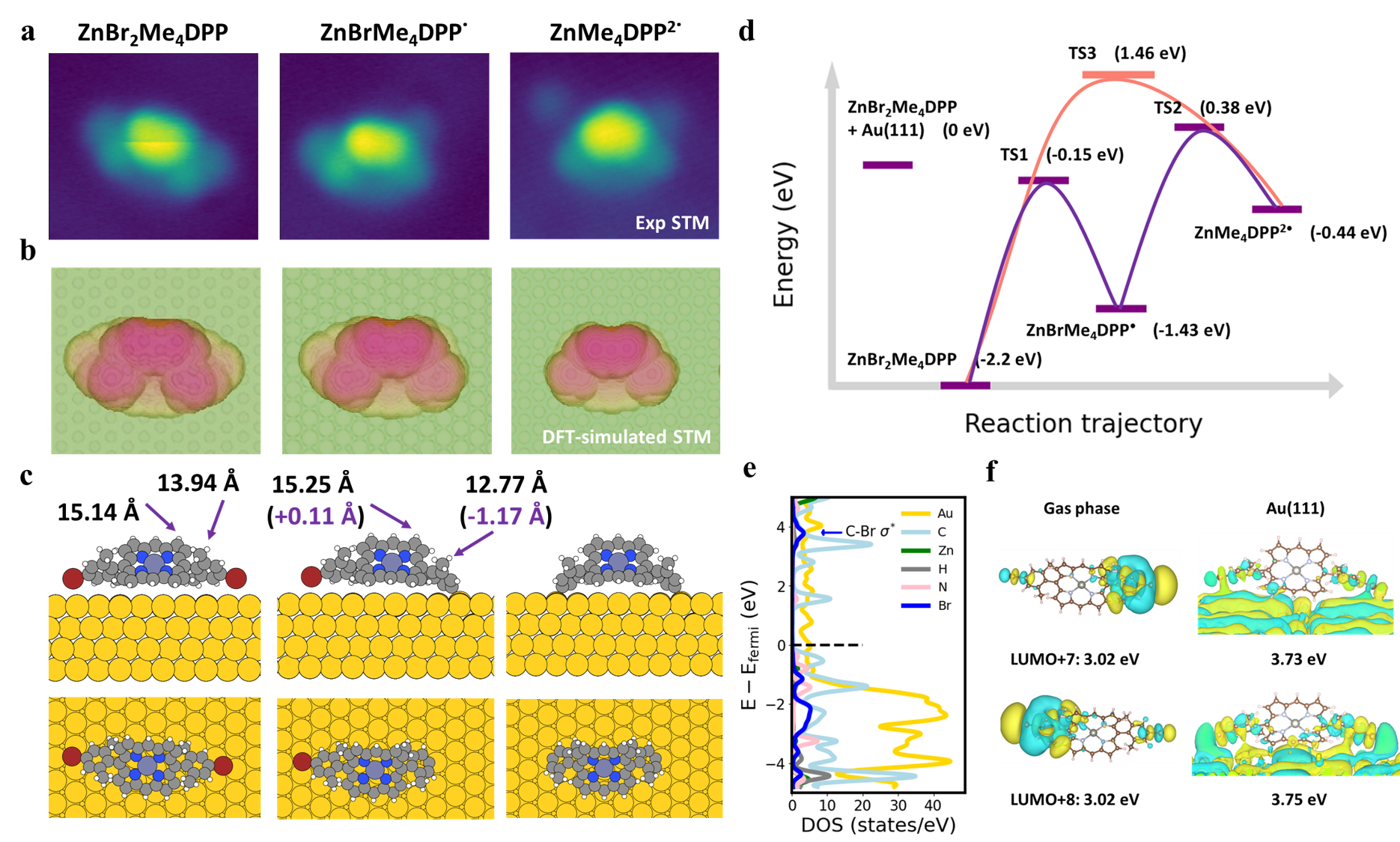}
\caption{\textbf{Interpretation of reaction.}
(a) STM images of \ch{ZnBr2Me4DPP}, \ch{ZnBrMe4DPP^{.}} and \ch{ZnMe4DPP^{2.}}. (b) Corresponding DFT-simulated STM images. (c) Corresponding 3D structures (top view and side view). (d) Corresponding energy diagrams from \ch{ZnBr2Me4DPP} to \ch{ZnBrMe4DPP^{.}} radical to \ch{ZnMe4DPP^{2.}} biradical. (e) Atom-Projected Density of States (PDOS) of \ch{ZnBr2Me4DPP} adsorbed on Au(111), all energies are relative to the Fermi level (-5.16 eV). The Au DOS are scaled by 0.05. (f) C--Br antibonding orbitals and states for the \ch{ZnBr2Me4DPP} molecule in the gas-phase as well as on surface.}
\label{reaction_mechanisms}
\end{figure}

In Fig. \ref{reaction_mechanisms}a, the STM images of pristine \ch{ZnBr2Me4DPP} (left), \ch{ZnBrMe4DPP^{.}} (center), \ch{ZnMe4DPP^{2.}} (right) are represented. Based on these, we explored the corresponding 3D structures of the \ch{ZnBrMe4DPP^{.}} or \ch{ZnMe4DPP^{2.}} products by DFT. When removing a Br atom from \ch{ZnBr2Me4DPP}, the adjacent C atom in the \ch{ZnBrMe4DPP^{.}} product tends to bind with an atom in the Au(111) surface, simultaneously losing its initial radical character, as corroborated by spin-polarized DFT computations. Similar effects are observed after removing the second Br atom, and the two terminals of the \ch{ZnMe4DPP^{2.}} product bind strongly with the surface, the whole molecule adopting an "arc-type" conformation on the surface. After inspecting the z-coordinates of specific atoms before and after dissociation, we note that some atoms are elevated, while others are lowered with respect to their initial positions (Fig. \ref{reaction_mechanisms}c), which could be reflected in the experimental topography (Fig. 3a, red dot is the initial height of tip). Furthermore, the energy diagrams (Fig. \ref{reaction_mechanisms}e) obtained by DFT calculations for the removals of two bromines from \ch{ZnBr2Me4DPP} in both concurrent and consecutive models show that the activation barrier energy for the first Br removal (2.05 eV) is close to the second Br removal (1.82 eV), while a higher activation energy barrier of 3.66 eV needs to be surmounted for the concurrent removal of two bromines. Therefore, the dissociation of two bromine atoms from \ch{ZnBr2Me4DPP} could occur sequentially or simultaneously during the dissociation procedure.

To gain further insight on the possible reaction mechanism leading to the C-Br bond dissociation, we analyzed the electronic structure of \ch{ZnBr2Me4DPP}. Highly localized C--Br antibonding orbitals were identified based on the feature in the PDOS corresponding to Br and C atoms at approximately 3.7 eV above the Fermi level (Fig. \ref{reaction_mechanisms}e). For the isolated gas-phase molecule, these correspond to LUMO+7 and LUMO+8. From visual inspection, the symmetries of the orbitals appear correct for a \textit{p-p} $\sigma$* bond, representing the antibonding counterpart for the single $\sigma$ bond between the C and Br atoms. Based on the agreement with the most successful bias voltage determined by RDL at 3.9 V, and the energies of the antibonding states, we surmise that one probable mechanism is dissociative electron attachment, where the tunneling electrons are supplied to the specific C--Br $\sigma$* orbitals. Subsequently, the charge is neutralized, which is accompanied by vibrational excitation of the C--Br bond. Furthermore, this provides us with a way to rationalize the observation that in some instances, both C--Br bonds are dissociated during manipulation, as the antibonding states are delocalized over both C--Br bonds, in contrast to the gas-phase molecule, where one end is more prominent. Despite this consistent model, we cannot fully rule out other mechanisms, as we observe successful dissociation with a wide range of parameters and tip positions. Considering that the computed barriers for dissociation are significantly lower than the $\sigma$* bond energies at 2.6 and 3.3 eV (Fig. \ref{reaction_mechanisms}e), respectively, another possible mechanism proceeds via tunneling into orbitals of lower energy, followed by the vibrational excitation. The resulting excess vibrational energy can quickly be redistributed into the other modes, including the one corresponding to C--Br bond vibration. This is consistent with the more stochastic nature (lower rate of success) of the dissociation observed at these lower bias values, as sufficient vibrational energy needs to be partitioned into these specific bonds. Furthermore, this also rationalizes why higher bias values induce more reactions in general, as more energy becomes available for reactions.

\newpage

\subsection{Signal classifier}

\begin{figure}[!th]%
\centering
\includegraphics[width=0.5\textwidth]
{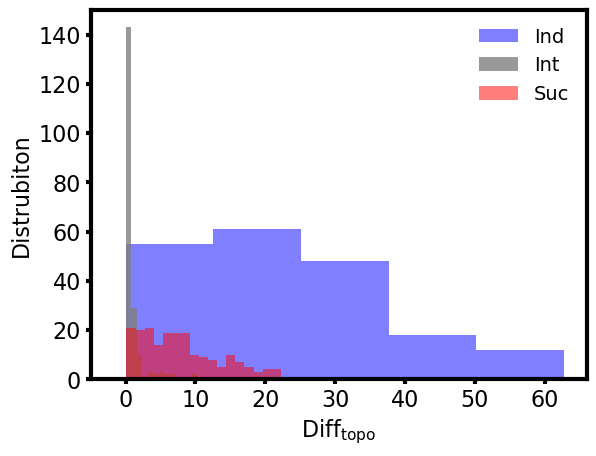}
\caption{Distribution of $Diff_{topo}$ during dissociation for three categories of products. 200 samples were obtained by a random selection from each category.}
\label{signal}
\end{figure}

\newpage
\subsection{Performance metrics of $\mathrm{M_{Target}}$}

\begin{figure}[!th]%
\centering
\includegraphics[width=0.9\textwidth]
{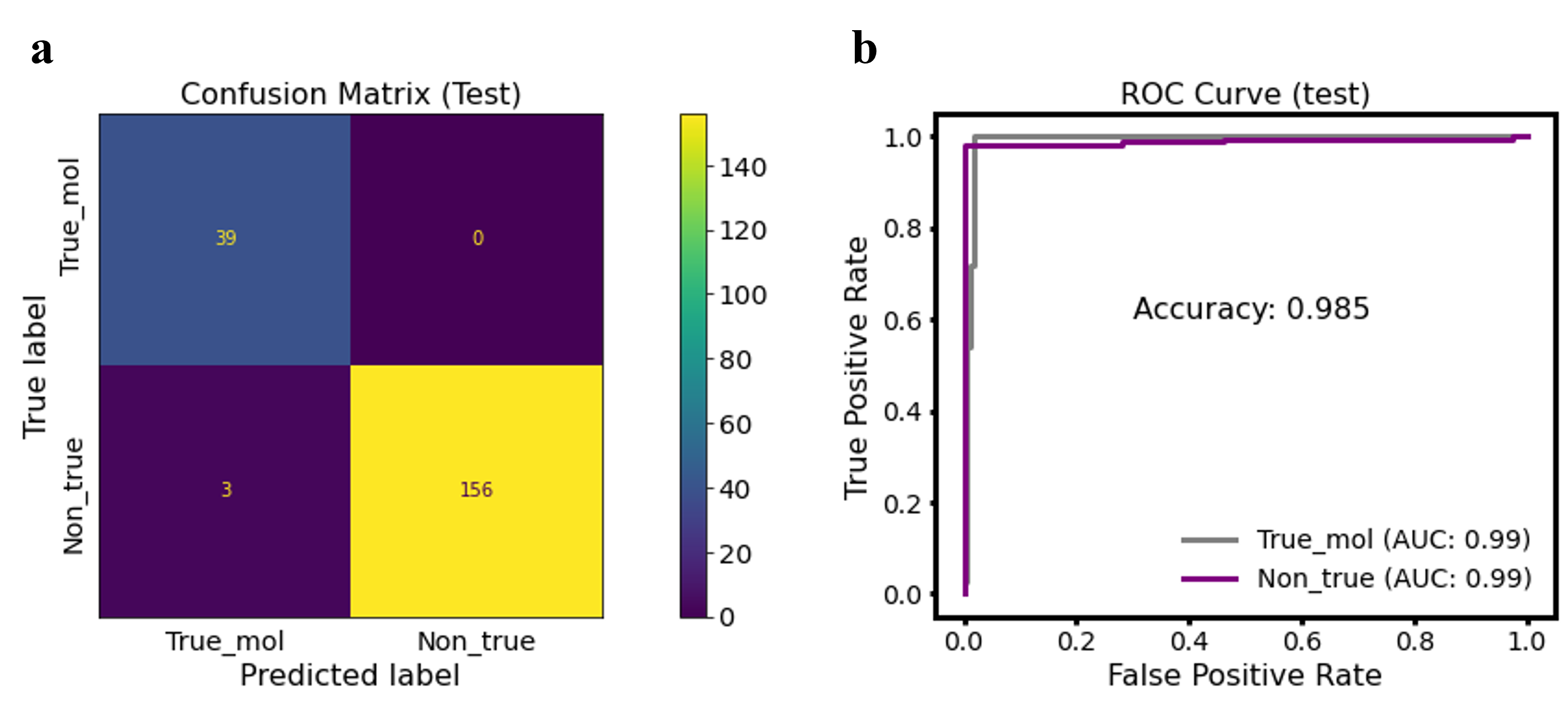}
\caption{Performance metrics of $\mathrm{M_{Target}}$ for distinguishing \ch{ZnBr2Me4DPP} or not when detecting targeted reactants, with 1389 samples in the training dataset and 198 samples in the test dataset.}
\label{target model}
\end{figure}

\begin{longtable}{@{}ccccccc@{}}
\caption{Performance metrics of $\mathrm{M_{Target}}$}\\ 
\begin{tabular}{ccccc}
\hline
\textbf{class} & \textbf{precision} & \textbf{recall} & \textbf{f1-score} & \textbf{support} \\ \hline
True\_mol      & 0.93               & 1.00            & 0.96              & 39               \\ 
Non\_true      & 1.00               & 0.98            & 0.99              & 159              \\ 
accuracy       & \multicolumn{3}{c}{0.98}                                 & 198              \\ 
macro avg      & 0.96               & 0.99            & 0.98              & 198              \\ 
weighted avg   & 0.99               & 0.98            & 0.99              & 198              \\ 

\end{tabular}

\end{longtable}

\newpage
\subsection{Performance metrics of $\mathrm{M_{Triple}}$}

\begin{figure}[!th]%
\centering
\includegraphics[width=0.9\textwidth]
{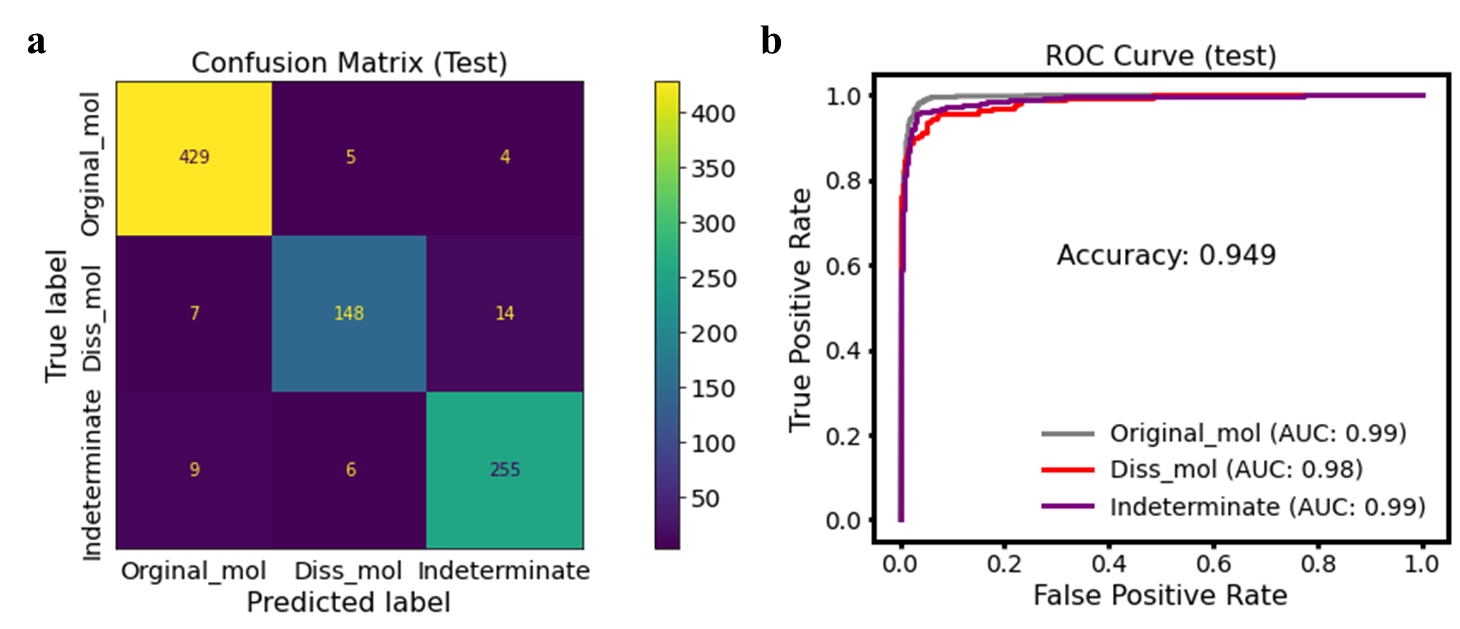}
\caption{Performance metrics of $\mathrm{M_{Triple}}$ for classifying \ch{ZnBr2Me4DPP}, dissociated molecules, indeterminate when evaluating products, with 3950 samples in the training dataset and 877 samples in the test dataset.}
\label{triple model}
\end{figure}

\begin{longtable}{@{}ccccccc@{}}
\caption{Performance metrics of $\mathrm{M_{Triple}}$}\\ 
\begin{tabular}{ccccc}
\hline
\textbf{class} & \textbf{precision} & \textbf{recall} & \textbf{f1-score} & \textbf{support} \\ \hline
Original\_mol        & 0.96               & 0.98            & 0.97              & 438               \\
Diss\_mol   & 0.93               & 0.88            & 0.90              & 169               \\
Indeterminate          & 0.93               & 0.94            & 0.94              & 270               \\
accuracy       & \multicolumn{3}{c}{0.95}                                 & 877              \\ 
macro avg      & 0.94               & 0.93            & 0.94              & 877              \\ 
weighted avg   & 0.95               & 0.95            & 0.95              & 877              \\         

\end{tabular}

\end{longtable}

\newpage
\subsection{Performance metrics of $\mathrm{M_{Ind}}$}

\begin{figure}[!th]%
\centering
\includegraphics[width=0.9\textwidth]
{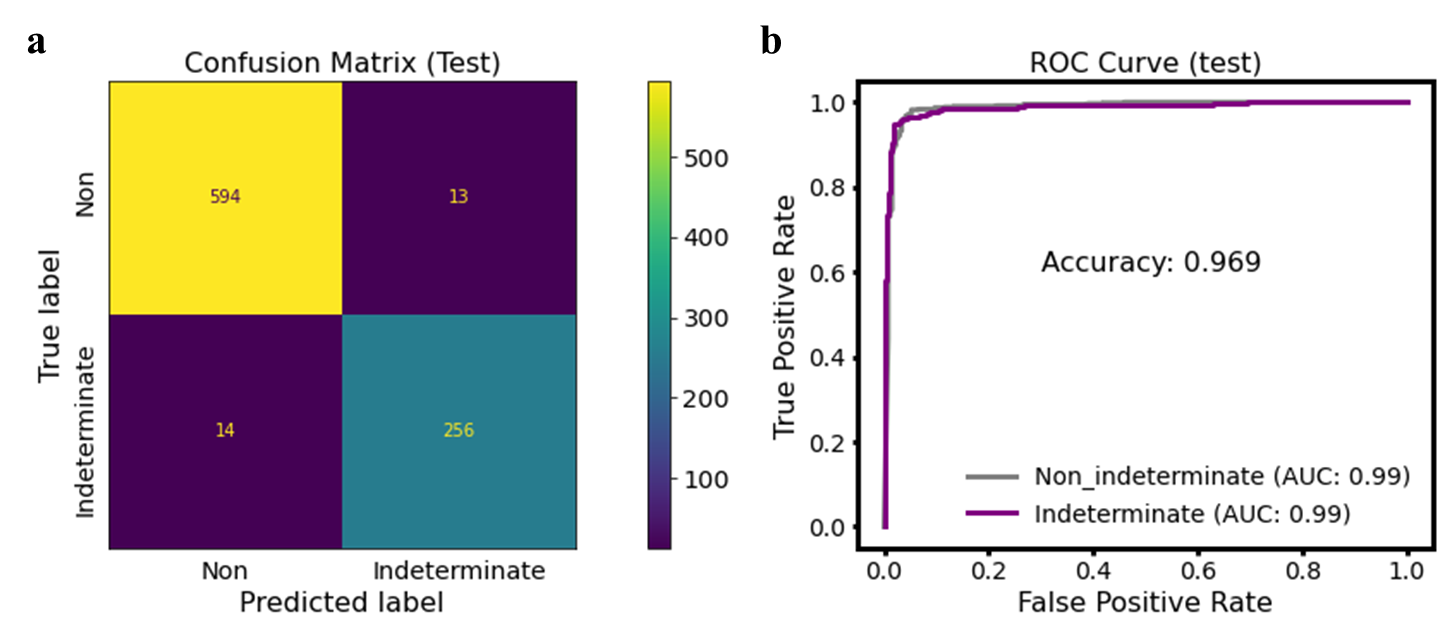}
\caption{Performance metrics of $\mathrm{M_{Ind}}$ for distinguishing indeterminate products among products, with 3950 samples in the training dataset and 877 samples in the test dataset.}
\label{ind model}
\end{figure}

\begin{longtable}{@{}ccccccc@{}}
\caption{Performance metrics of $\mathrm{M_{Ind}}$}\\ 
\begin{tabular}{ccccc}
\hline
\textbf{class} & \textbf{precision} & \textbf{recall} & \textbf{f1-score} & \textbf{support} \\ \hline
Non\_indeterminate     & 0.98               & 0.98            & 0.98              & 607              \\
Indeterminate          & 0.95               & 0.95            & 0.95              & 270              \\
accuracy       & \multicolumn{3}{c}{0.97}                                 & 877              \\
macro avg      & 0.96               & 0.96            & 0.96              & 877              \\
weighted avg   & 0.97               & 0.97            & 0.97              & 877              \\ 

\end{tabular}

\end{longtable}

\newpage
\subsection{Performance metrics of $\bf{M_{Diss}}$}

\begin{figure}[!th]%
\centering
\includegraphics[width=0.9\textwidth]
{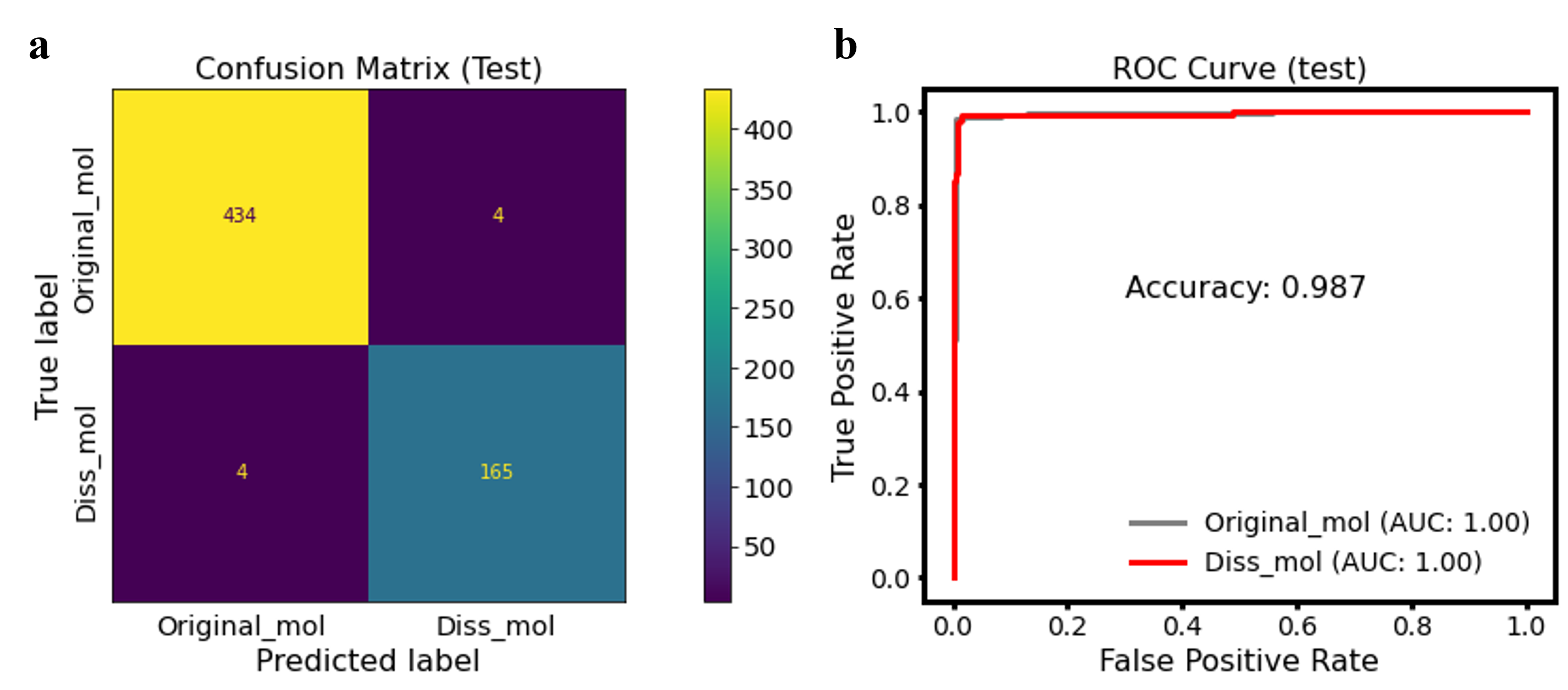}
\caption{Performance metrics of $\mathrm{M_{Diss}}$ for distinguishing \ch{ZnBr2Me4DPP} and dissociated molecules when evaluating products, with 2764 samples in the training dataset and 607 samples in the test dataset.}
\label{diss model}
\end{figure}

\begin{longtable}{@{}ccccccc@{}}
\caption{Performance metrics of $\mathrm{M_{Diss}}$}\\ 
\begin{tabular}{ccccc}
\hline
\textbf{class} & \textbf{precision} & \textbf{recall} & \textbf{f1-score} & \textbf{support} \\ \hline
Original\_mol  & 0.99               & 0.99            & 0.99              & 438               \\
Diss\_mol      & 0.98               & 0.98            & 0.98              & 169               \\
accuracy       & \multicolumn{3}{c}{0.99}                                 & 607              \\
macro avg      & 0.98               & 0.98            & 0.98              & 607               \\
weighted avg   & 0.99               & 0.99            & 0.99              & 607              \\

\end{tabular}

\end{longtable}

\newpage
\section{Decision-making}
\subsection{Random action at different bias patterns}

\begin{figure}[!th]%
\centering
\includegraphics[width=\textwidth]
{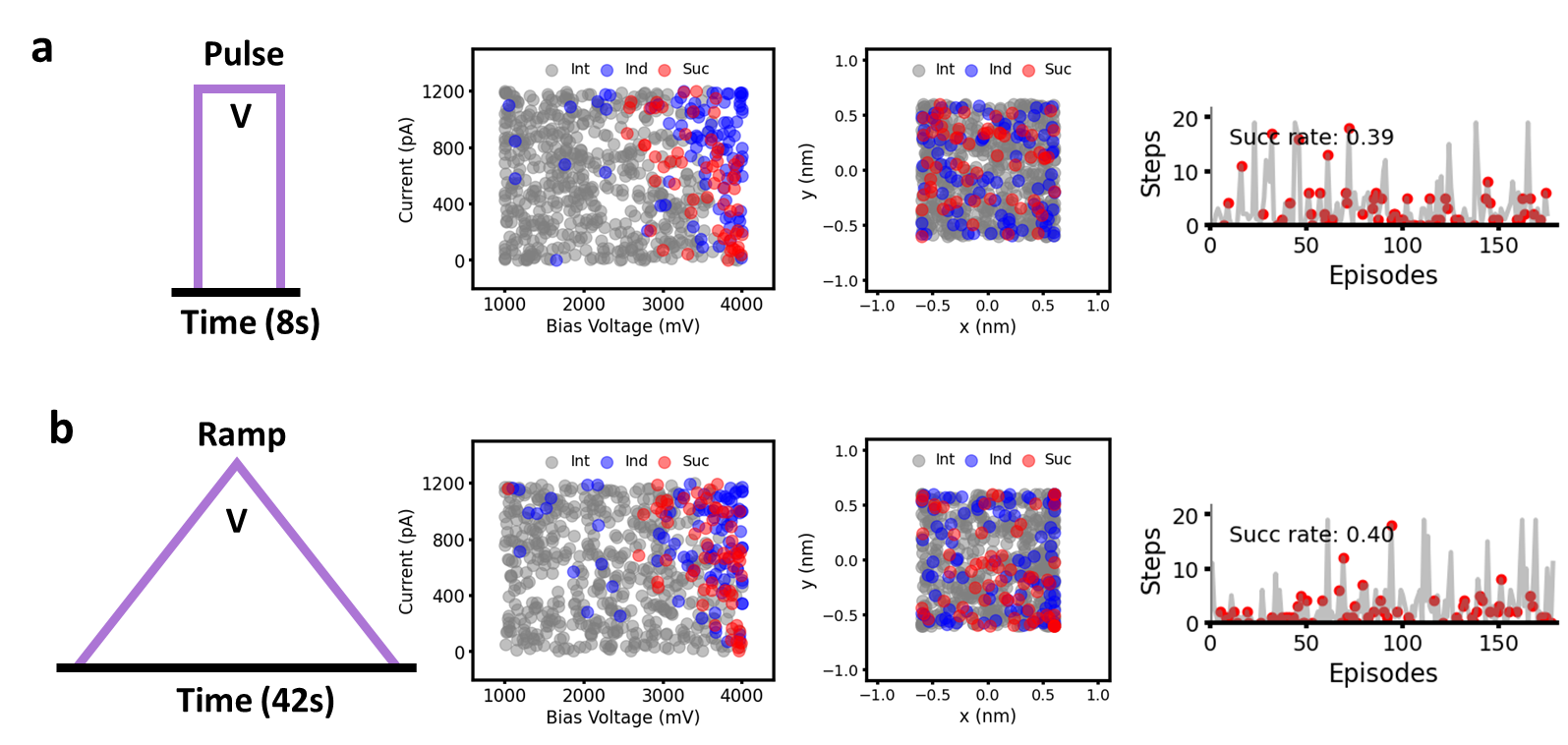}
\caption{\textbf{Performance of randomly generating actions at two types of bias patterns with tip positions referred to the centre point of
the ellipse constrained to within 0.6 nm along x (or y) axis.} (a) 8 s pulse pattern. (b) 42 s ramp pattern. The first column illustrates the bias voltage patterns, the second column indicates the distribution of Suc (red dots), Ind (blue dots), and Int species (grey dots) upon varying the current and the voltage, the third column shows the corresponding distribution upon varying the tip position (tip$_x$ and tip$_y$), the fourth column represents the evolution of dissociation steps over multiple episodes, where red points symbolize successful dissociation.}
\label{pulse and ramp}
\end{figure}

\newpage
\subsection{Contrast pattern measurement}

\begin{figure}[!th]%
\centering
\includegraphics[width=0.4\textwidth]
{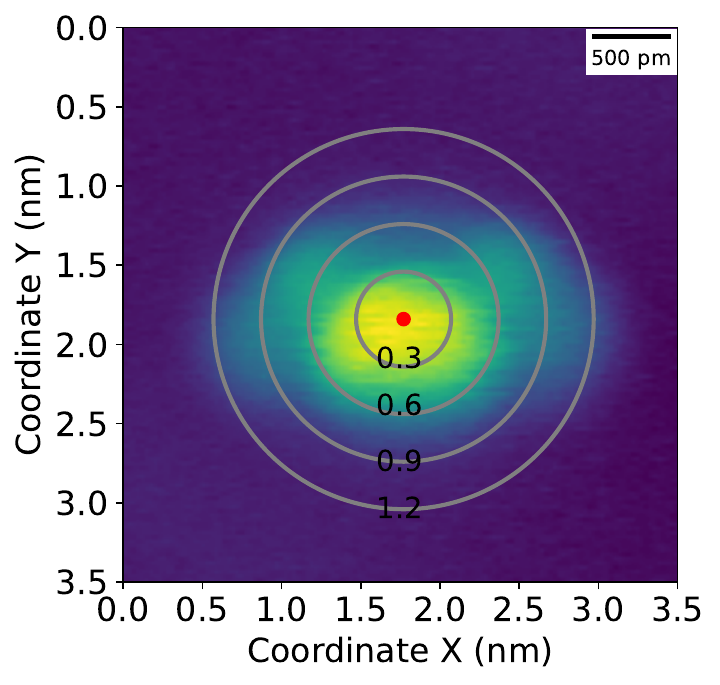}
\caption{Schematic illustration of the distance of each part of \ch{ZnBr2Me4DPP} within a pattern approximated as an ellipse, referenced to its center point.}
\label{protrusion}
\end{figure}

\newpage
\subsection{Random action with fixed tip position}

\begin{figure}[!th]%
\centering
\includegraphics[width=0.9\textwidth]
{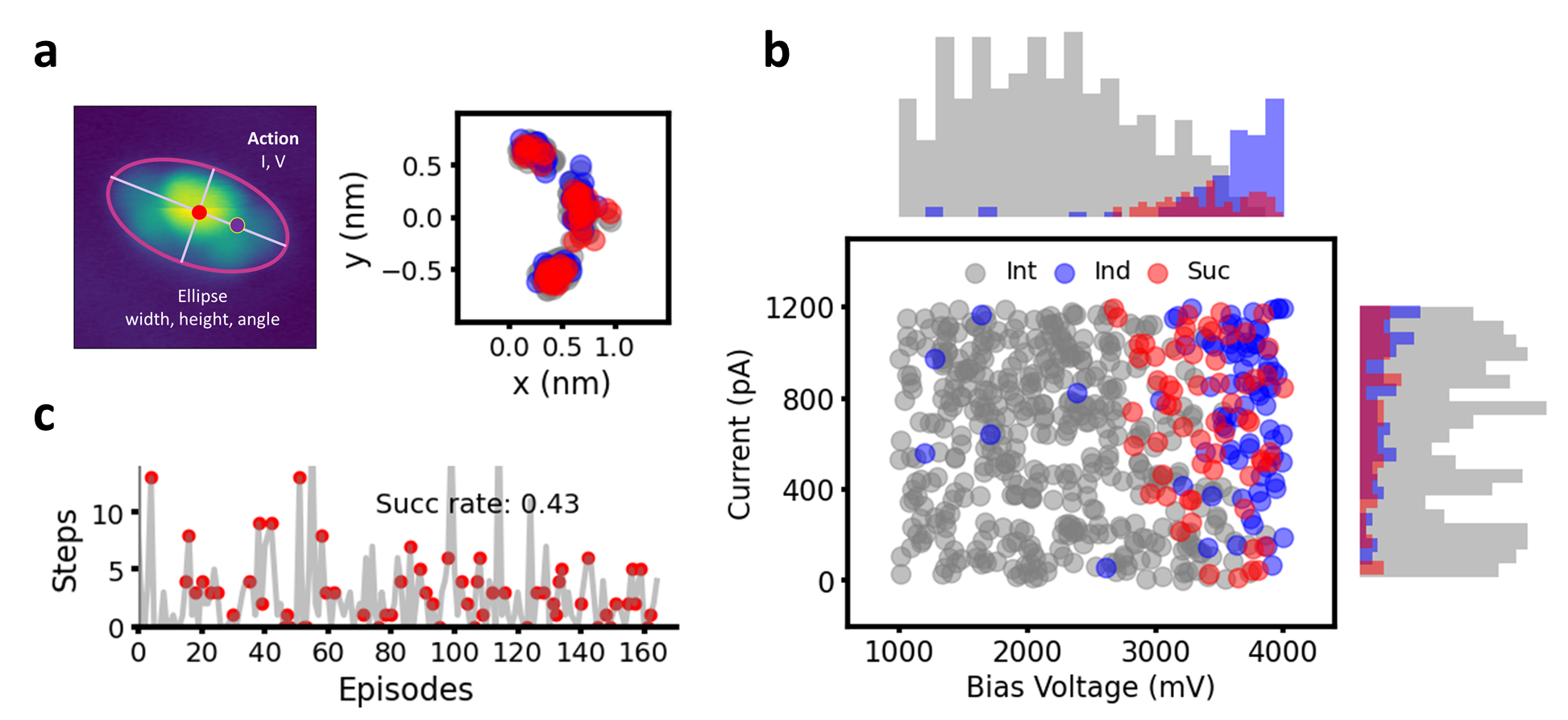}
\caption{\textbf{Performance of random action at a fixed tip position.} (a) Left panel: the pattern corresponding to a molecule is approximated as an ellipse. Tip position referred to the center point of the ellipse. Right panel: the distribution of tip position in this test. (b) Distribution of voltages and currents. Three categories of products (Int, Ind, Suc) under the voltage-current pairs are marked by grey, red and blue. (c) The evolution of dissociation steps over multiple episodes, red points indicate successful dissociation. The specific tip position is set based on Equation 6.}
\label{random and fix}
\end{figure}

\newpage

\subsection{Random selection from successful dissociation parameters}

\begin{figure}[!th]%
\centering
\includegraphics[width=0.9\textwidth]
{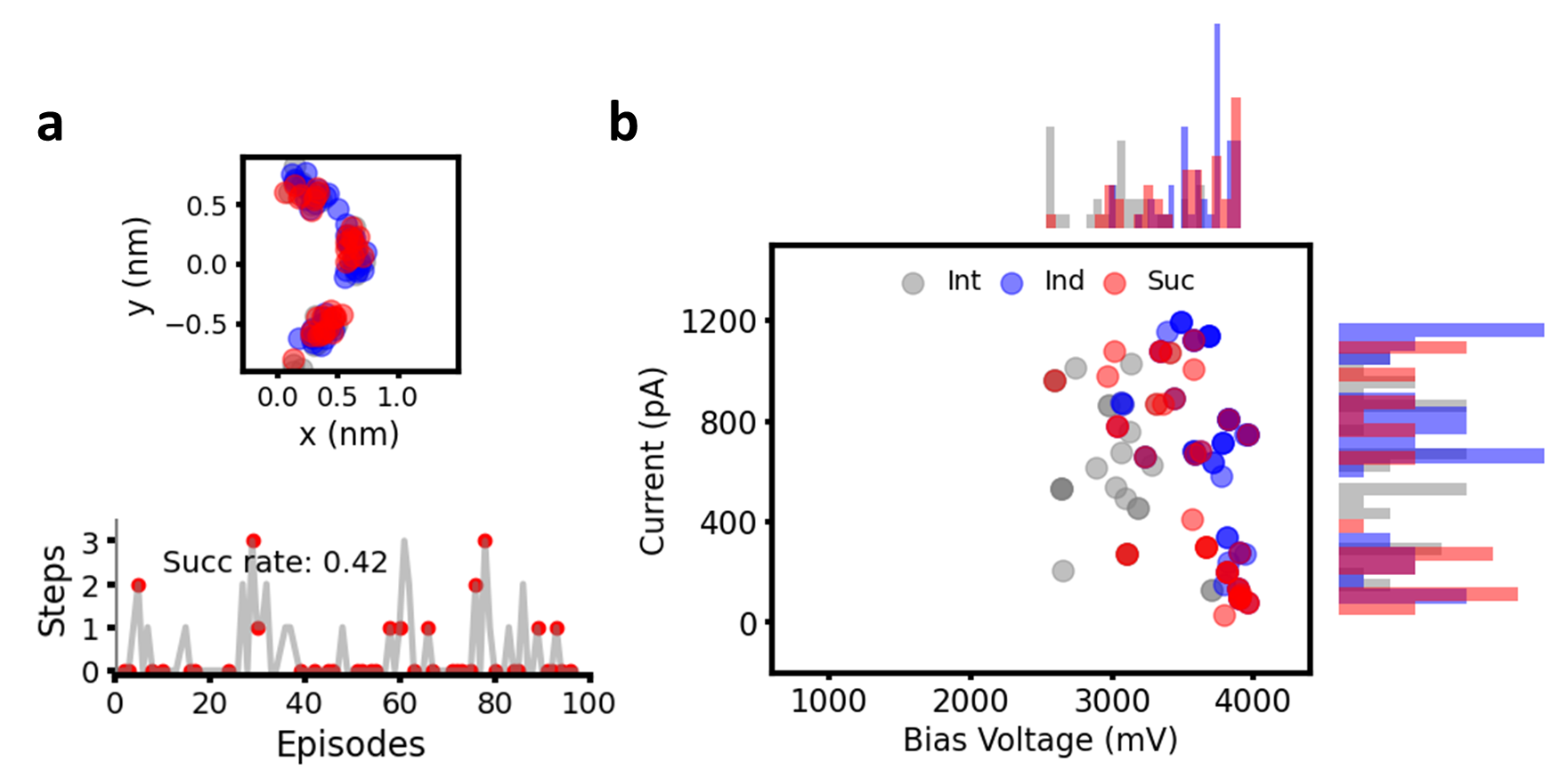}
\caption{\textbf{Performance of randomly selecting actions from successful dissociation parameters.}(a) Upper panel: Distribution of the tip position, Lower pane: Evolution of dissociation steps over multiple episodes, red points indicate
successful dissociation. (b) Distribution of voltages and currents.}
\label{random and succ}
\end{figure}

\subsection{Dissociation parameters from the DRL model}

\begin{figure}[!th]%
\centering
\includegraphics[width=0.9\textwidth]
{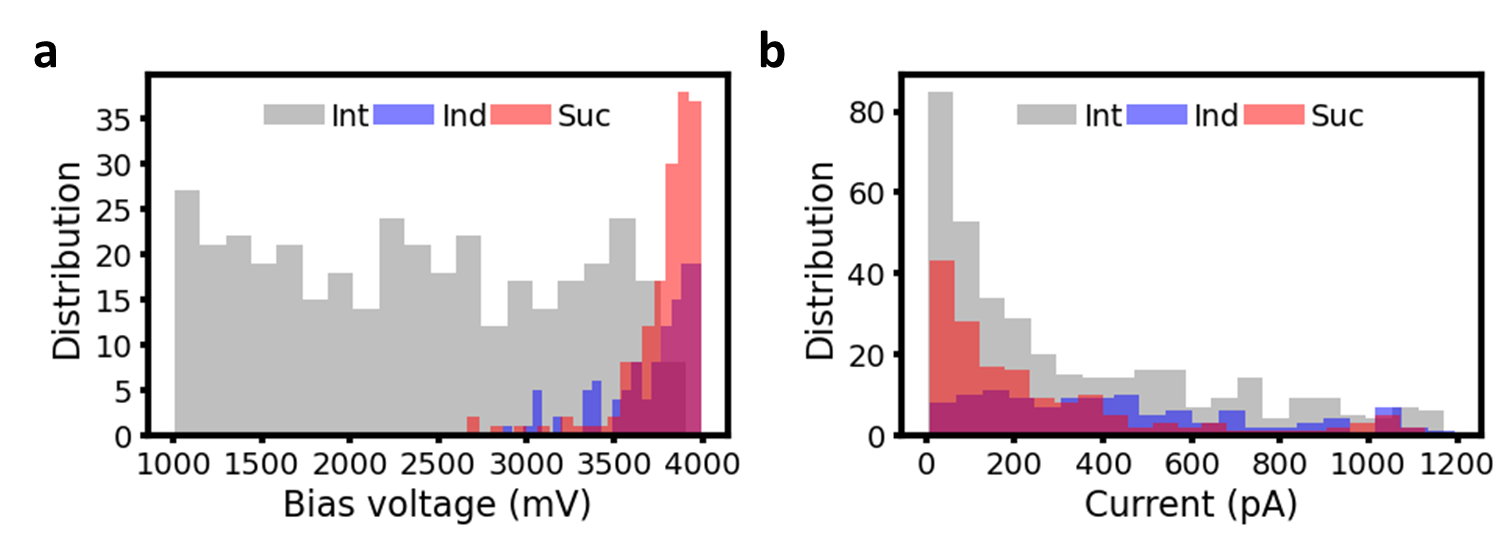}
\caption{Distribution of bias voltages (a) and currents (b) under three categories of products (Int, Ind and Suc), marked by grey, red and blue, in the DRL training}
\label{sac_distribution}
\end{figure}

\newpage

\subsection{Tip conditioning region}

\begin{figure}[!th]%
\centering
\includegraphics[width=0.4\textwidth]
{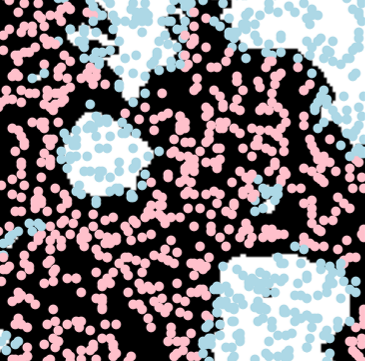}
\caption{Schematic illustration of our algorithm detecting blank regions and selecting tip position from specific regions for conditioning tips.}
\label{tip condition}
\end{figure}

\newpage
\section{Synthesis and characterization}
\subsection{Materials and Methods}

Chemicals and solvents were purchased from commercial suppliers (Aldrich, BLDpharm, Thermo Scientific Chemicals and Scharlab) and used without further purification. All dry solvents were freshly distilled under argon over an appropriate drying agent before use. Column chromatography was carried out on silica gel VWR-60 (40-63 µm). Analytical TLC was performed on aluminum sheets precoated with silica gel 60 F-254 from Merck. 
\textsuperscript{1}H- and \textsuperscript{13}C-NMR spectra were recorded with a Bruker XRD-500 (500 MHz) instrument at room temperature (25 ºC). Chemical shifts values ($\delta$) are referred to the corresponding deuterated solvent (\ch{CDCl3}). 
UV/vis experiments were carried out by using quartz cells with a 1 cm optical path length in a JASCO-V660 UV-vis spectrophotometer.
MALDI-TOF MS spectra were obtained in a Bruker ULTRAFLEX III (MALDI-TOF/TOF) spectrometer.

\begin{figure}[!th]%
\centering
\includegraphics[width=0.9\textwidth]
{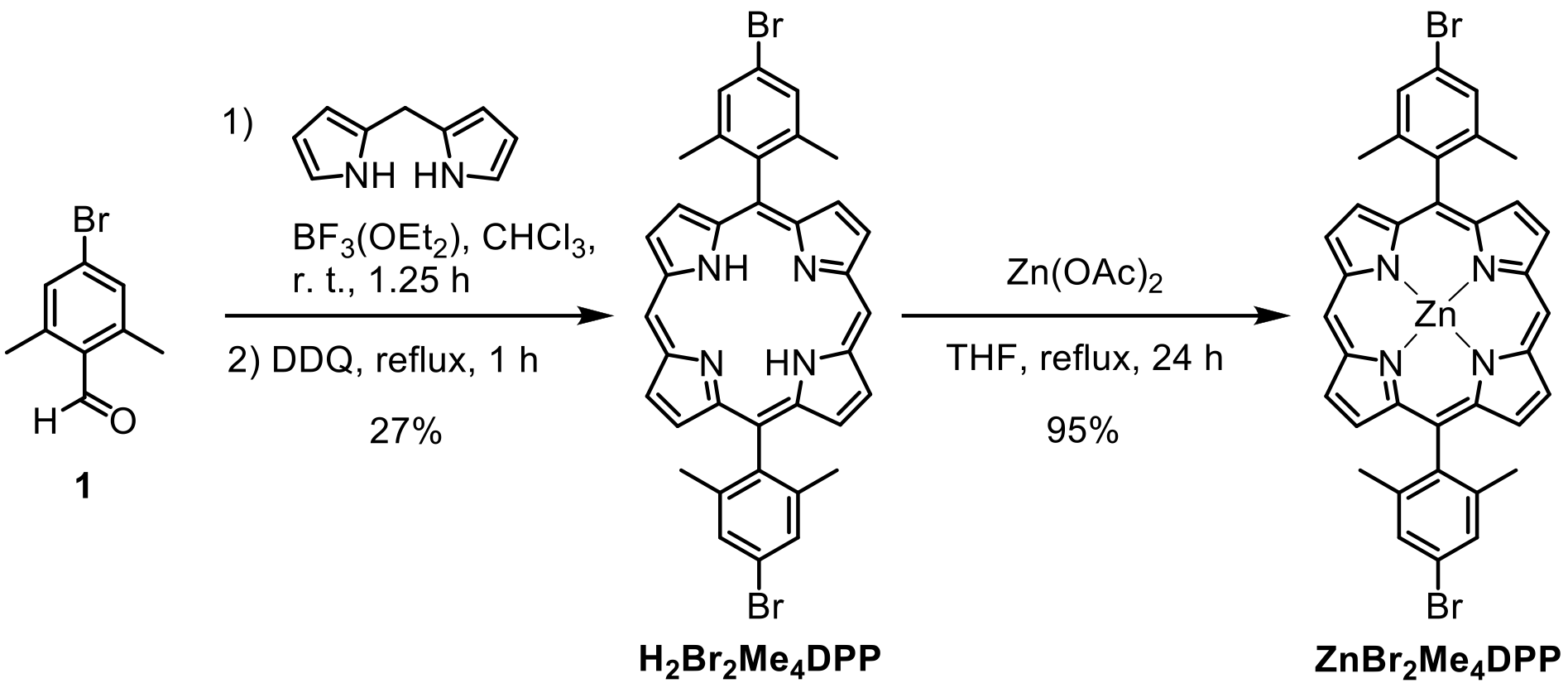}
\caption{Synthetic route towards \ch{ZnBr2Me4DPP}.}
\label{reaction route}
\end{figure}

\subsection{Synthesis of precursors \ch{H2Br2Me4DPP}}

\ch{H2Br2Me4DPP} was synthesized using a modified procedure from Zhao and co-workers\cite{zhao2023quantum}.
meso-H-dipyrromethane (533 mg, 2.5 mmol, 1 eq.) and 2,6-dimethyl-4-bromobenzaldehyde (365 mg, 2.5 mmol, 1 eq.) were loaded in a flame-dried 500 mL round-bottom flask. \ch{CHCl3} (250 mL) and dry ethanol (1.87 mL) were added to the flask and the mixture was vigorously degassed with argon for 40 minutes under stirring and in the dark. Then, \ch{BF3(OEt2)} (0.1 mL) was added under argon, whereupon the mixture turned dark red. The mixture was stirred in the dark for 80 minutes at room temperature, then DDQ (908 mg, 4 mmol, 1.6 eq.) was added in one batch and the mixture was heated to reflux for 1 h. After that, the mixture was allowed to cool down to room temperature, triethylamine (2 mL) was added and the solution stirred for 5 minutes. After this time, the solvent was removed under reduced pressure and the resulting crude purified by column chromatography using DCM/n-heptane (3:2) as eluent. From the column, a red-coloured fraction was isolated (Rf = 0.68) which upon removal of the solvent under reduced pressure afforded a solid which was suspended in n-pentane, sonicated, filtered, washed with n-pentane (4 mL), and dried under vacuum to yield \ch{H2Br2Me4DPP} as a dark-violet powder (225 mg, 27\%). \textsuperscript{1}H-NMR (500 MHz, \ch{CDCl3}): $\delta$ = 10.26 (s, 2H), 9.37 (d, J = 4.6 Hz, 4H), 8.87 (d, J = 4.6 Hz, 4H), 7.69 (s, 4H), 1.86 (s, 12H), –3.10 (s, 2H); \textsuperscript{13}C NMR (125 MHz, \ch{CDCl3}): $\delta$ = 142.0, 139.7, 132.4, 130.1, 129.9, 122.4, 115.9, 105.1, 21.7; MALDI-TOF MS (DCTB matrix): m/z (\% intensity) = 674.1 – 681.1 m/z $\mathrm{[M]^{+}}$ (100\%); UV/vis (\ch{CHCl3}): $\lambda$max (log $\epsilon$) = 405 (5.32), 500 (4.21), 576 (3.72).

\clearpage
\subsubsection{\textsuperscript{1}H-NMR}

\begin{figure}[!th]%
\centering
\includegraphics[width=0.73\textwidth]
{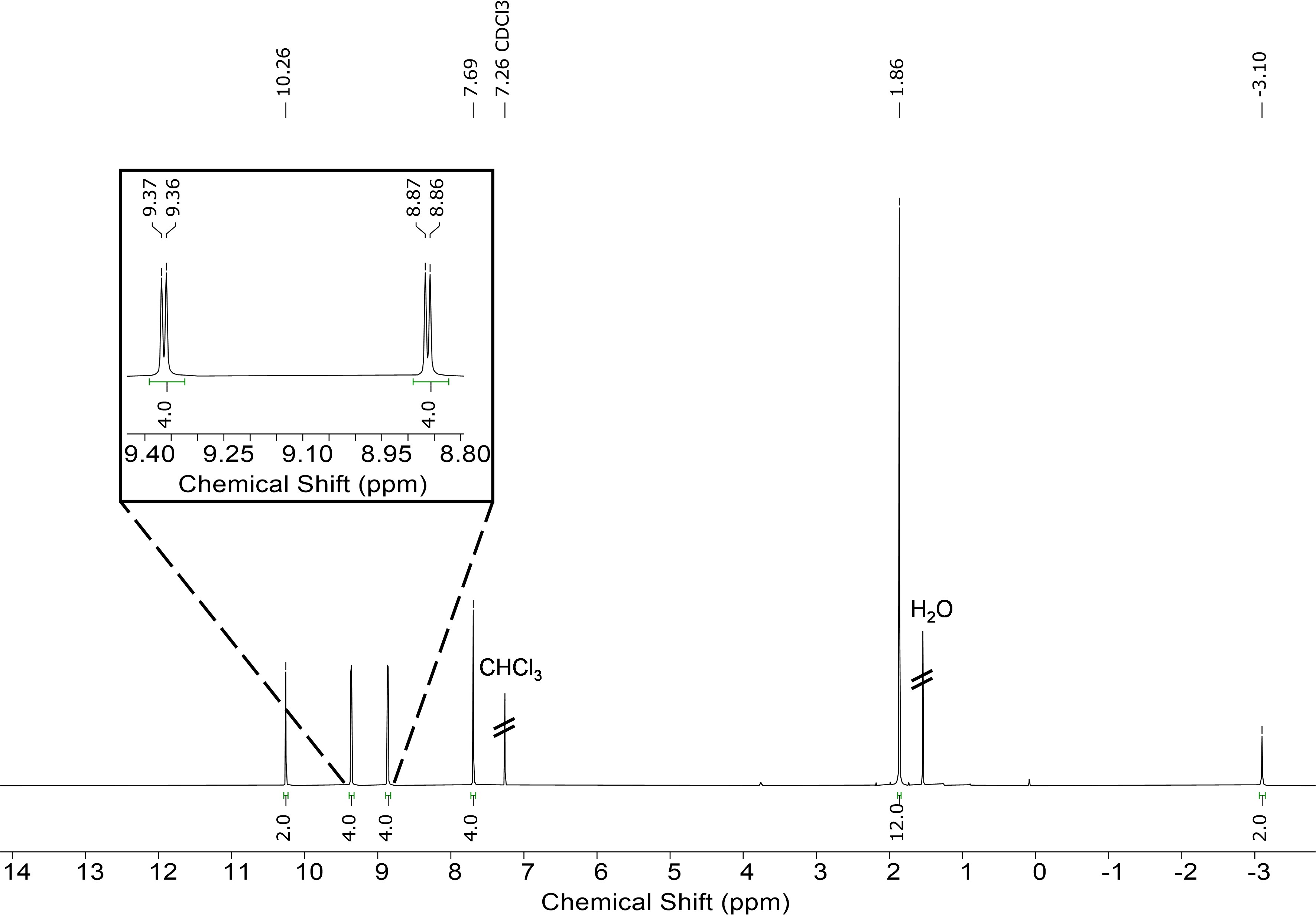}
\caption{\textsuperscript{1}H-NMR spectrum of \ch{H2Br2Me4DPP} in \ch{CDCl3} ($ \rotatebox{45}{=} $ = residual solvent signals).}
\label{HNMR-H2}
\end{figure}

\subsubsection{\textsuperscript{13}C-NMR}

\begin{figure}[!th]%
\centering
\includegraphics[width=0.73\textwidth]
{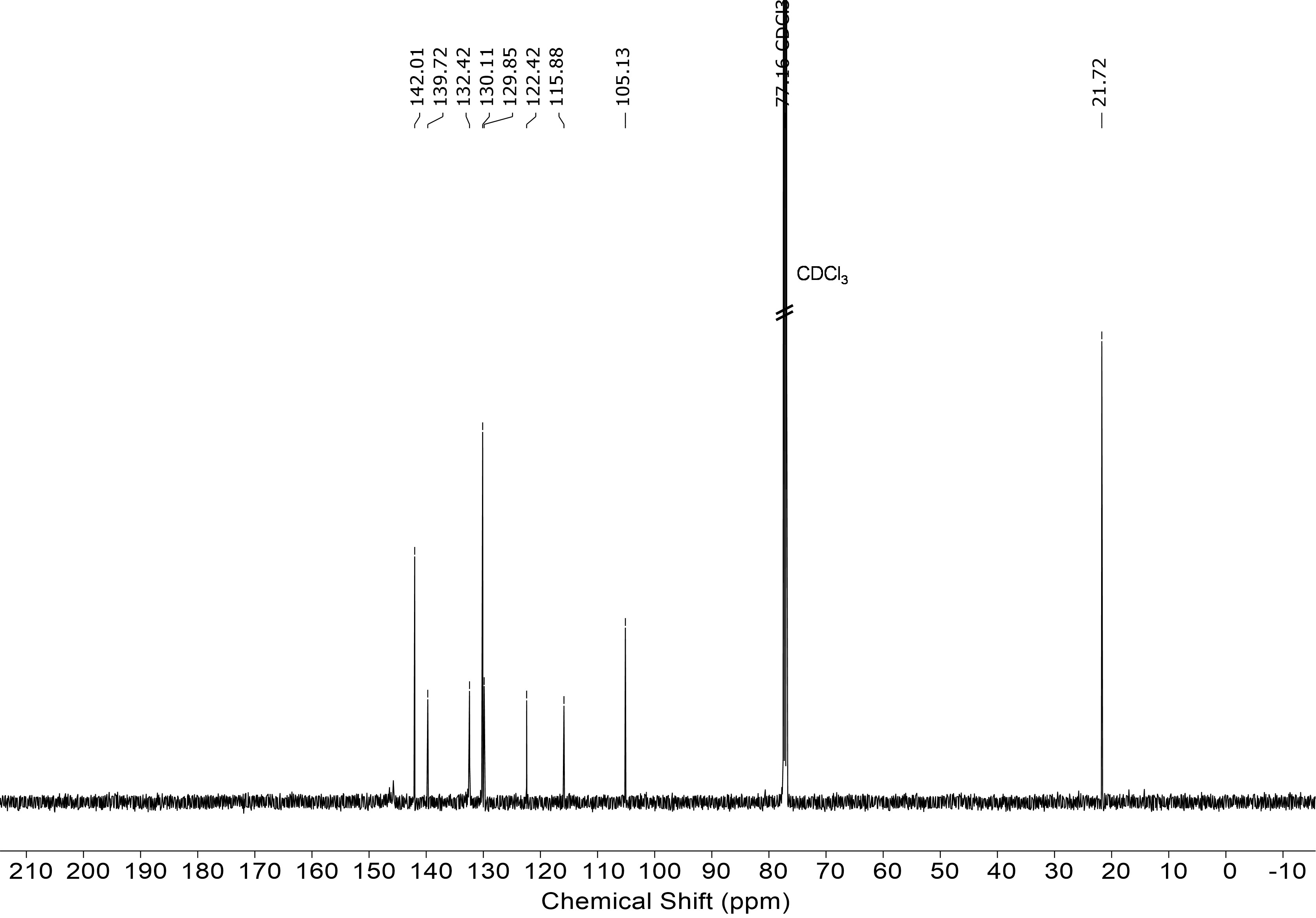}
\caption{\textsuperscript{13}C-NMR spectrum of \ch{H2Br2Me4DPP} in \ch{CDCl3} ($ \rotatebox{45}{=} $ = residual solvent signals).}
\label{CNMR-H2}
\end{figure}

\newpage
\subsubsection{MS}

\begin{figure}[!th]%
\centering
\includegraphics[width=0.73\textwidth]
{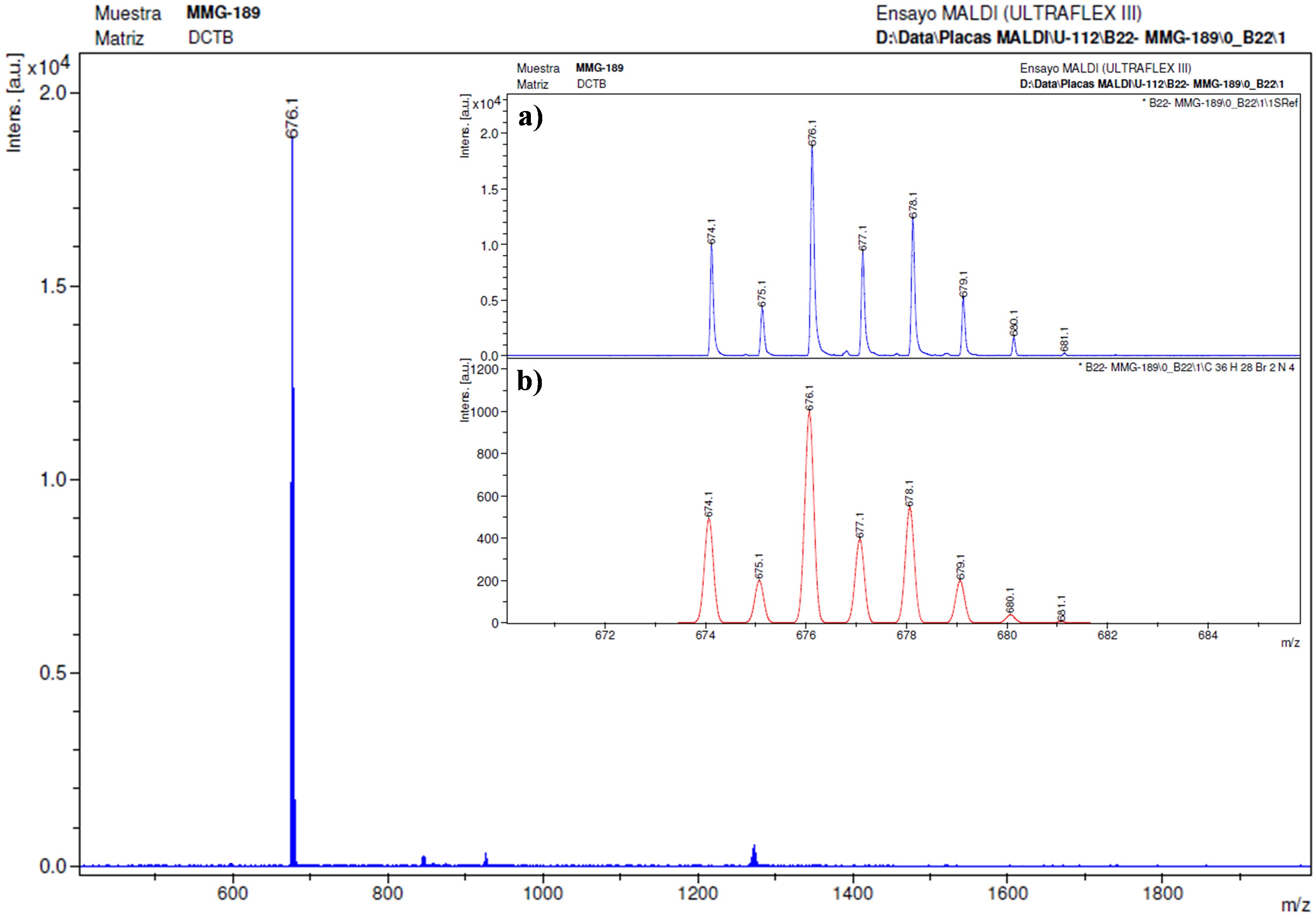}
\caption{MALDI-TOF mass spectrum of \ch{H2Br2Me4DPP}. Inset: (a) experimental isotopic resolution of the MALDI-TOF main peak at 676.1 m/z. (b) Calculated isotopic pattern for \ch{H2Br2Me4DPP}.}
\label{MS-H2}
\end{figure}

\subsubsection{UV/vis}

\begin{figure}[!th]%
\centering
\includegraphics[width=0.65\textwidth]
{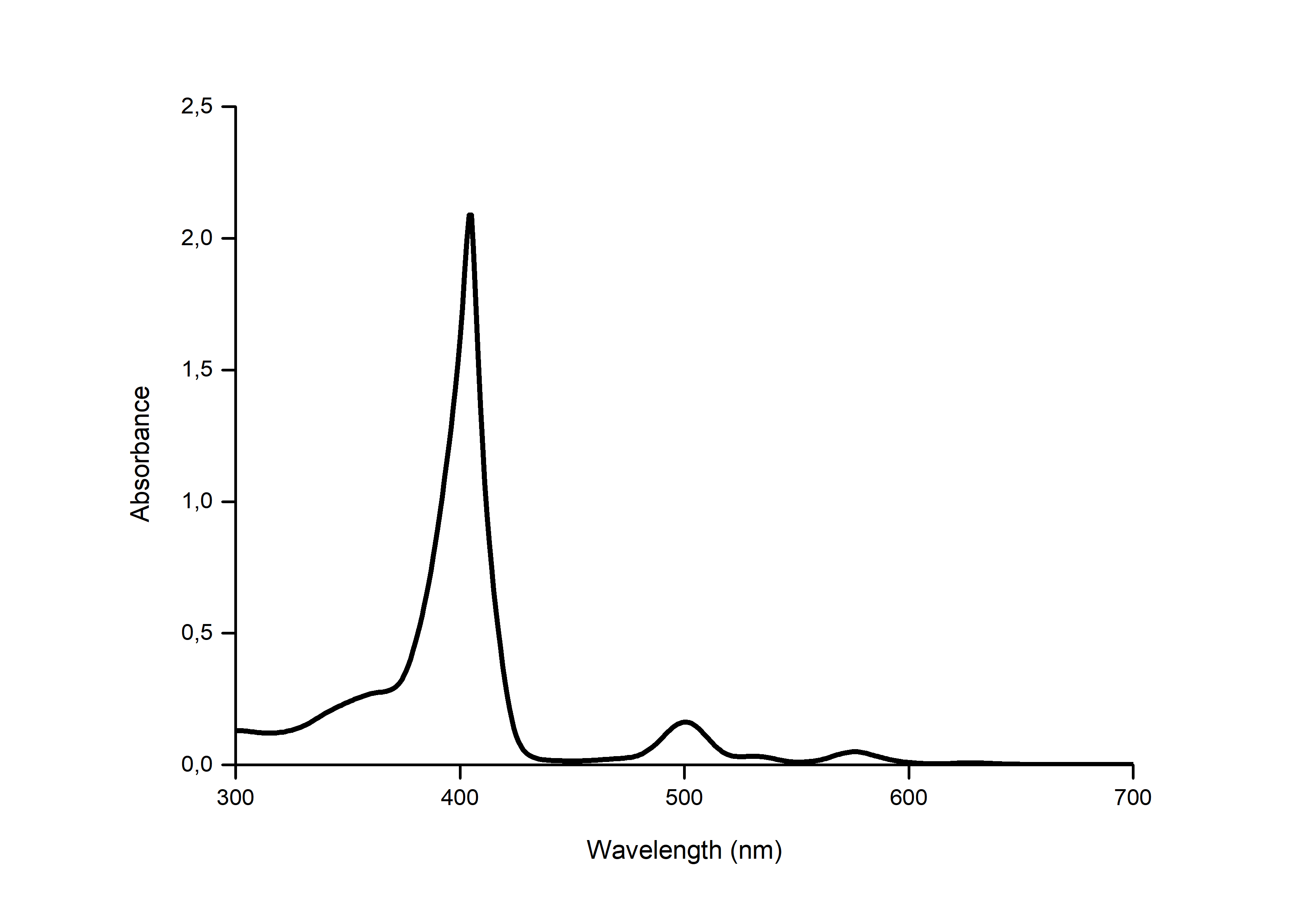}
\caption{UV/vis spectrum of \ch{H2Br2Me4DPP} in \ch{CHCl3} (conc. = 10 µM).}
\label{UV-H2}
\end{figure}

\clearpage

\subsection{Synthesis of \ch{ZnBr2Me4DPP}}

\ch{H2Br2Me4DPP} (20 mg, 0.03 mmol) and \ch{Zn(OAc)2} (54 mg, 0.30 mmol, 10 eq.) were loaded in a flame-dried 10 mL Schlenk tube and three cycles of vacuum/argon backfilling were applied. Then, dry THF (5 mL) was added, and the mixture was heated to reflux while stirring under argon for 24 h. After this time, the mixture was then allowed to cool to room temperature, quenched with water (10 mL) and extracted with DCM (3 × 15 mL). The combined organic layers were washed with water (25 mL) and brine (25 mL), dried over anhydrous \ch{MgSO4}, filtered and the filtrate dried under reduced pressure. The resulting crude was subjected to a short column chromatography using DCM/n-heptane (3:2) as eluent, isolating a red-coloured fraction (Rf = 0.61) which solvent was removed under reduced pressure. The resulting solid was suspended in n-pentane, sonicated, filtered, washed with n-pentane (3 × 2 mL), and dried under vacuum to yield \ch{ZnBr2Me4DPP} as a purple powder (21 mg, 95\%). \textsuperscript{1}H-NMR (500 MHz, \ch{CDCl3}): $\delta$ = 10.28 (s, 2H), 9.42 (d, J = 4.5 Hz, 4H), 8.94 (d, J = 4.5 Hz, 4H), 7.68 (s, 4H), 1.83 (s, 12H); \textsuperscript{13}C-NMR (125 MHz, \ch{CDCl3}): $\delta$ = 149.8, 149.5, 141.9, 140.9, 132.8, 131.1, 129.9, 122.1, 116.9, 106.1, 21.7; MALDI-TOF MS (DCTB matrix): m/z (\% intensity) = 736.0 – 746.0 m/z $\mathrm{[M]^{+}}$ (100\%); UV/vis (\ch{CHCl3}): $\lambda$max (log $\epsilon$) = 410 (5.34), 540 (4.22). 

\newpage

\subsubsection{\textsuperscript{1}H-NMR}

\begin{figure}[!th]%
\centering
\includegraphics[width=0.73\textwidth]
{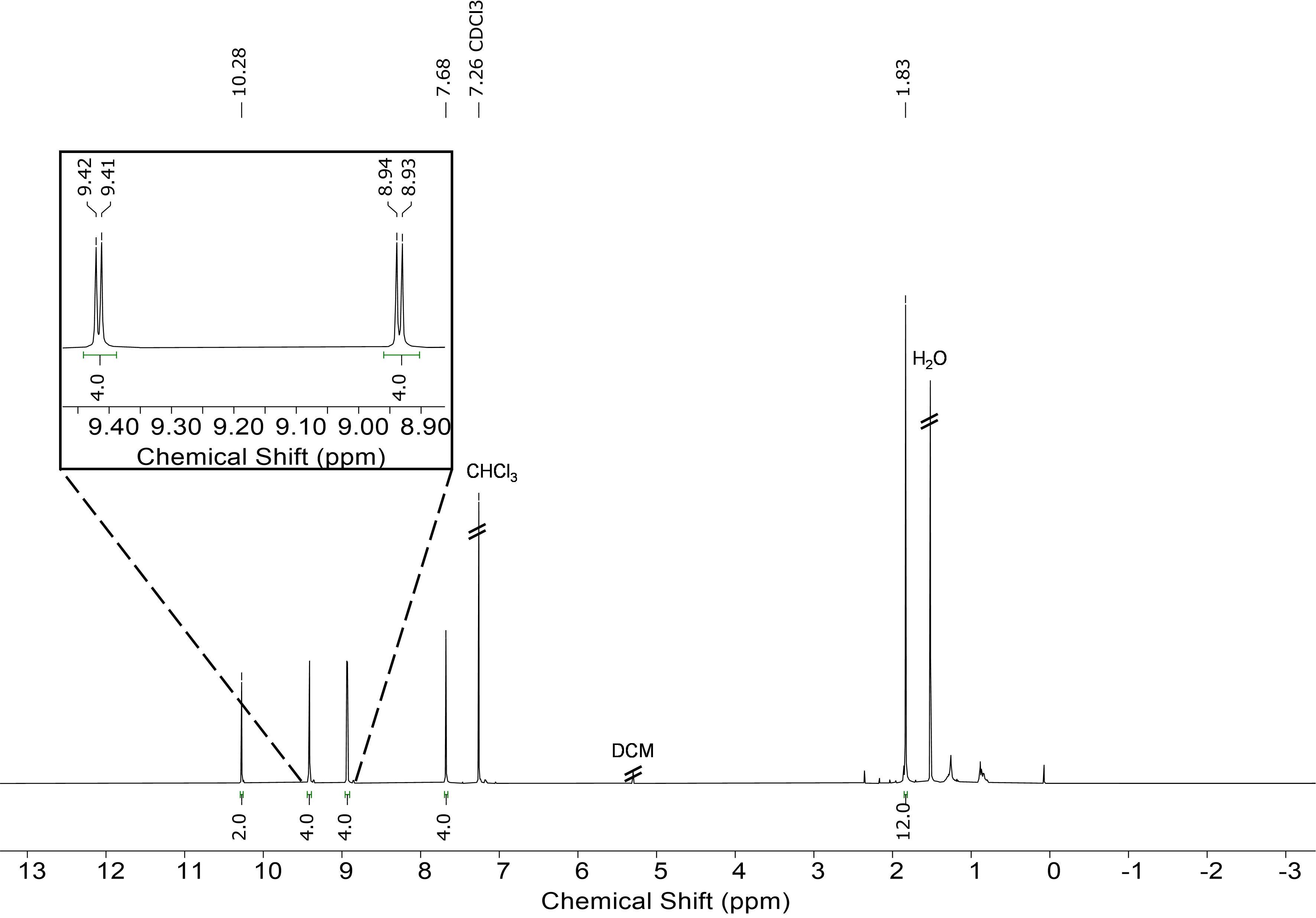}
\caption{\textsuperscript{1}H-NMR spectrum of \ch{ZnBr2Me4DPP} in \ch{CDCl3} ($ \rotatebox{45}{=} $ = residual solvent signals).}
\label{HNMR-Zn}
\end{figure}

\subsubsection{\textsuperscript{13}C-NMR}

\begin{figure}[!th]%
\centering
\includegraphics[width=0.73\textwidth]
{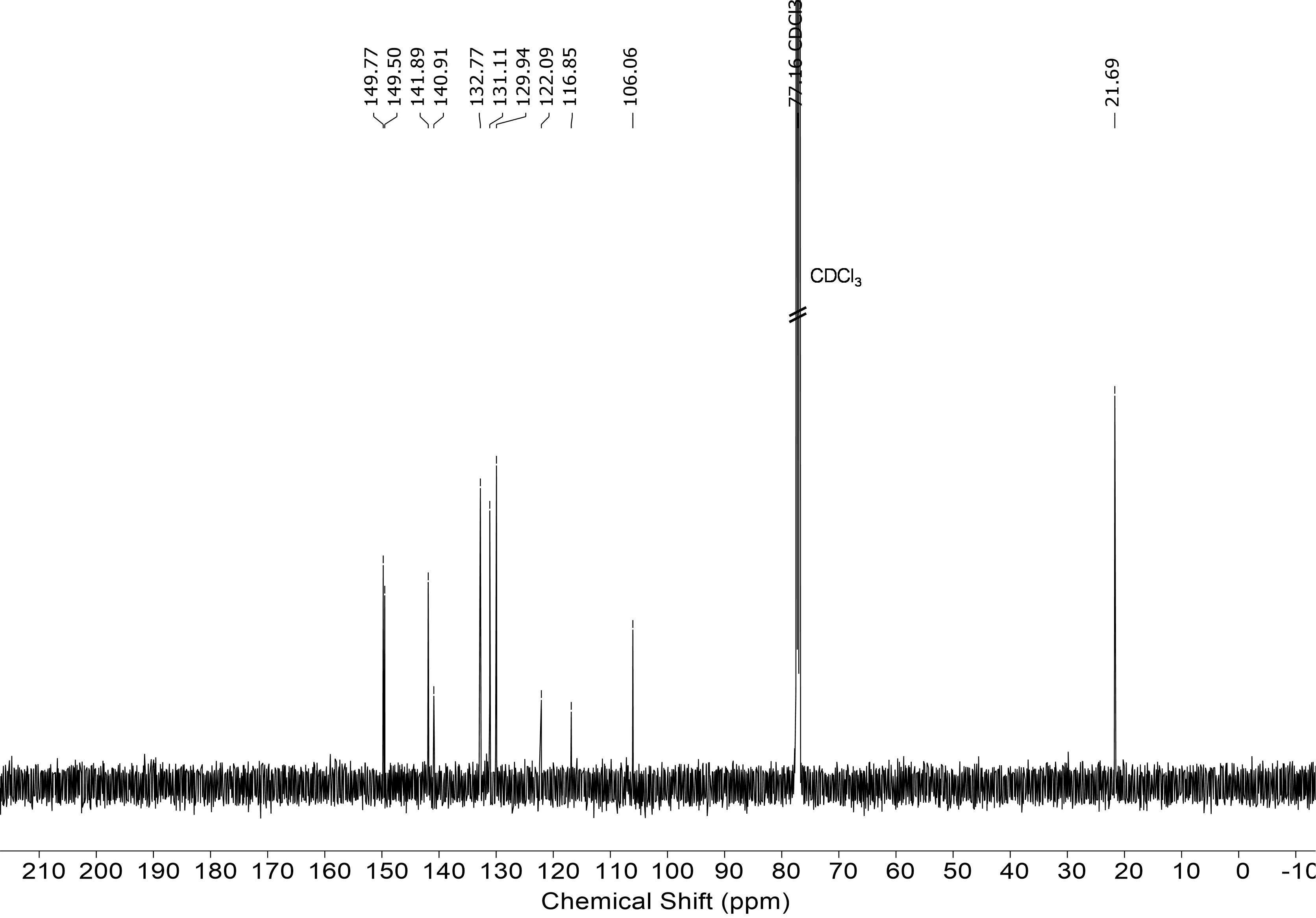}
\caption{\textsuperscript{13}C-NMR spectrum of \ch{ZnBr2Me4DPP} in \ch{CDCl3} ($ \rotatebox{45}{=} $ = residual solvent signals).}
\label{CNMR-Zn}
\end{figure}

\subsubsection{MS}

\begin{figure}[!th]%
\centering
\includegraphics[width=0.73\textwidth]
{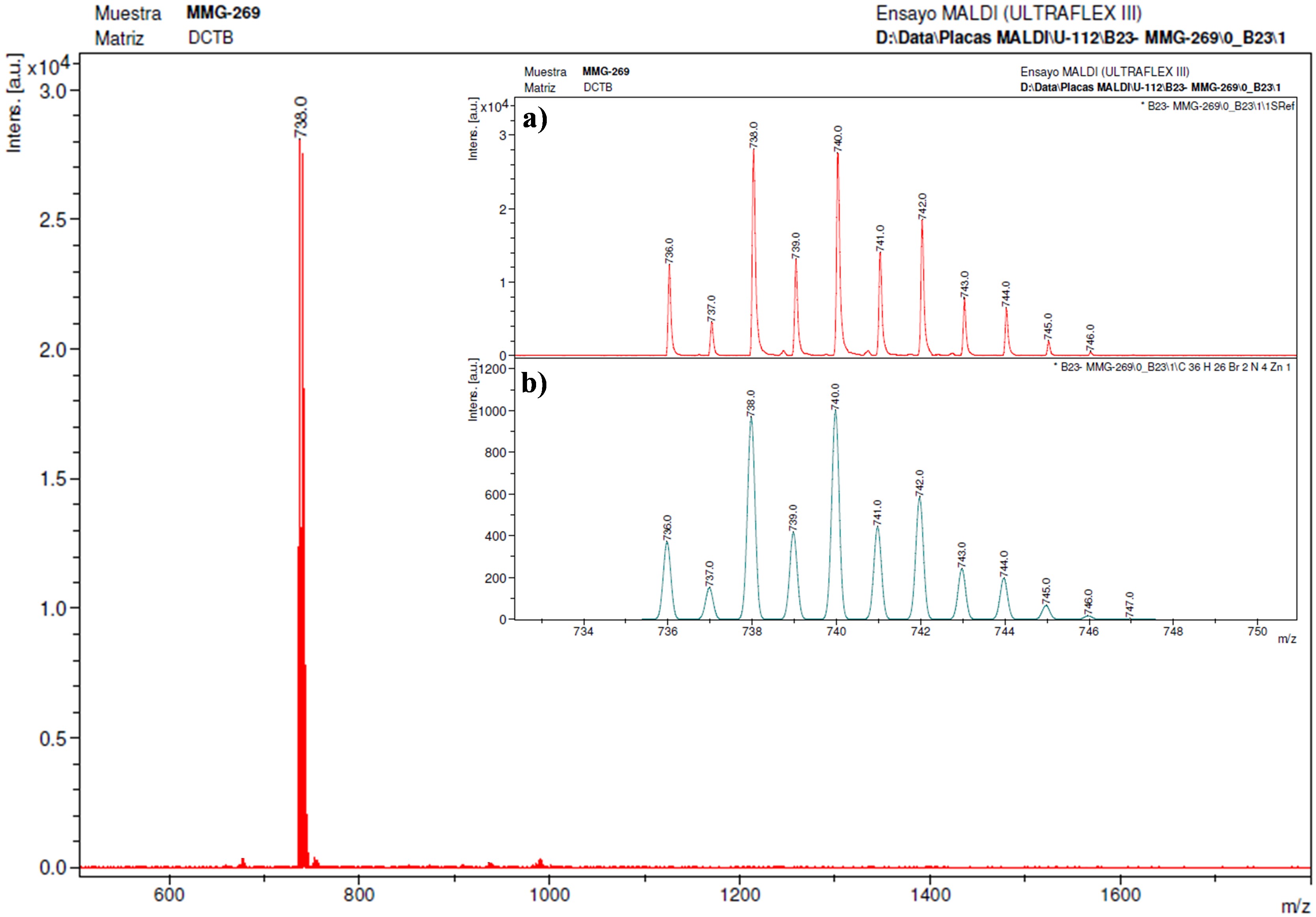}
\caption{MALDI-TOF mass spectrum of \ch{ZnBr2Me4DPP}. Inset: (a) experimental isotopic resolution of the MALDI-TOF main peak at 738.0 m/z. (b) Calculated isotopic pattern for \ch{ZnBr2Me4DPP}.}
\label{MS-Zn}
\end{figure}

\subsubsection{UV/vis}

\begin{figure}[!th]%
\centering
\includegraphics[width=0.65\textwidth]
{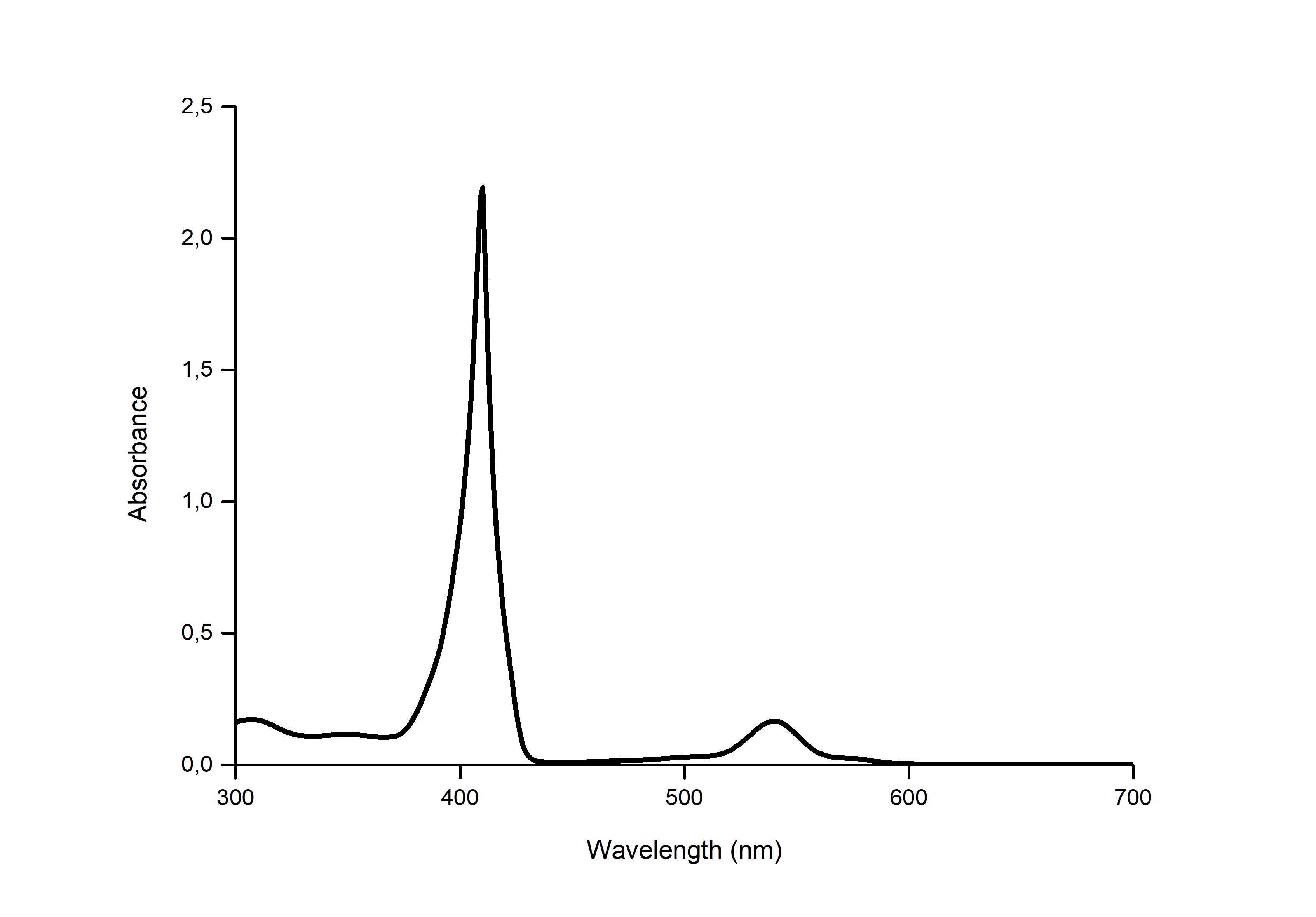}
\caption{UV/vis spectrum of \ch{ZnBr2Me4DPP} in \ch{CHCl3} (conc. = 10 µM).}
\label{UV-Zn}
\end{figure}

\bibliography{achemso-demo}